\documentclass[apj]{emulateapj}

\slugcomment{Draft of \today}

\shorttitle{ULXs in the IR with Spitzer}
\shortauthors{Lau et al.}
\usepackage{natbib}
\usepackage{amsmath}
\usepackage{graphicx}
\usepackage[section] {placeins}
\usepackage{hyperref}
\usepackage{ulem}
\usepackage{longtable}
\usepackage{rotating}
\usepackage{hyperref}

\newcommand{\beq}{\begin{equation}}
\newcommand{\eeq}{\end{equation}}

\newcommand{\spitzer}{\textit{Spitzer}}

\newcommand{\Heida}{H14}
\newcommand{\Lopez}{L17}
\newcommand{\NULX}{96}
\newcommand{\NIRULX}{12}
\newcommand{\ODULX}{16}
\newcommand{\NDULX}{68}
\newcommand{\NIRMULX}{12}

\begin{document}

\title{Uncovering Red and Dusty Ultraluminous X-ray Sources with Spitzer}

\author{Ryan M. Lau\altaffilmark{1},\altaffilmark{2}}
\author{Marianne Heida\altaffilmark{2}}
\author{Dominic J. Walton\altaffilmark{3}}
\author{Mansi M. Kasliwal\altaffilmark{2}}
\author{Scott M. Adams\altaffilmark{2}}
\author{Ann Marie Cody\altaffilmark{4}}
\author{Kishalay De\altaffilmark{2}}
\author{Robert D. Gehrz\altaffilmark{5}}
\author{Felix F{\"u}rst\altaffilmark{6}}
\author{Jacob E. Jencson\altaffilmark{2}}
\author{Jamie A. Kennea\altaffilmark{7}}
\author{Frank Masci\altaffilmark{8}}

\altaffiltext{1}{Institute of Space \& Astronautical Science, Japan Aerospace Exploration Agency, 3-1-1 Yoshinodai, Chuo-ku, Sagamihara, Kanagawa 252-5210, Japan}
\altaffiltext{2}{California Institute of Technology, Pasadena, CA 91125, USA}
\altaffiltext{3}{Institute of Astronomy, Madingley Road, CB3 0HA Cambridge, United Kingdom}

\altaffiltext{4}{NASA Ames Research Center, Moffett Field, CA 94035, USA}
\altaffiltext{5}{Minnesota Institute for Astrophysics, School of Physics and Astronomy, 116 Church Street, S. E., University of Minnesota, Minneapolis, MN 55455, USA}
\altaffiltext{6}{Dept. of Astronomy \& Astrophysics, Pennsylvania State University, University Park, PA 16802 USA}
\altaffiltext{7}{European Space Astronomy Centre (ESA/ESAC), Operations Department, Villanueva de la Ca\~{n}ada (Madrid), Spain}
\altaffiltext{8}{Infrared Processing and Analysis Center, California Institute of Technology, M/S 100-22, Pasadena, CA 91125, USA}

\begin{abstract}

We present a mid-infrared (IR) sample study of nearby ultraluminous X-ray sources (ULXs) using multi-epoch observations with the \textit{Infrared Array Camera} (IRAC) on the \textit{Spitzer Space Telescope}. \spitzer/IRAC observations taken after 2014 were obtained as part of the Spitzer Infrared Intensive Transients Survey (SPIRITS). Our sample includes \NULX~ULXs located within 10 Mpc. Of the \NULX~ULXs, \NIRULX~have candidate counterparts consistent with absolute mid-IR magnitudes of supergiants, and \ODULX~counterparts exceeded the mid-IR brightness of single supergiants and are thus more consistent with star clusters or non-ULX background active galactic nuclei (AGN). The supergiant candidate counterparts exhibit a bi-modal color distribution in a Spitzer/IRAC color-magnitude diagram, where ``red' and ``blue" ULXs fall in IRAC colors $[3.6] - [4.5]\sim0.7$ and $[3.6] - [4.5]\sim0.0$, respectively. The mid-IR colors and absolute magnitudes of 4 ``red" and 5 ``blue" ULXs are consistent with that of supergiant B[e] (sgB[e]) and red supergiant (RSG) stars, respectively. While ``blue",
RSG-like mid-IR ULX counterparts likely host RSG mass donors, we propose the ``red" counterparts are ULXs exhibiting the ``B[e] phenomenon'' rather than hosts of sgB[e] mass donors.  We show that the mid-IR excess from the ``red" ULXs is likely due to thermal emission from circumstellar or circumbinary dust. Using dust as a probe for total mass, we estimate mass-loss rates of $\dot{M}\sim1\times10^{-4}$ M$_\odot$ yr$^{-1}$ in dust-forming outflows of red ULXs.
Based on the transient mid-IR behavior and its relatively flat spectral index, $\alpha=-0.19\pm0.1$, we suggest that the mid-IR emission from Holmberg IX X-1 originates from a variable jet.

\end{abstract}

\maketitle

\section{Introduction}

An Ultraluminous X-ray source (ULXs) is an off-nuclear point-like source that exhibits an X-ray luminosity $>10^{39}$ erg s$^{-1}$, which exceeds the Eddington limit for a 10 M$_\odot$ black hole (Feng \& Soria 2011; Kaaret et al. 2017). Despite their comparable luminosities, ULXs are not active galactic nuclei (AGN) powered by super-massive black holes (Fabbiano 2006). The extreme luminosities exhibited by ULXs are therefore believed to be driven by accretion onto a compact object in a close binary (i.e. X-ray binaries). However, the accretion rates required to power the ULX emission far exceed the Eddington limit for white dwarfs, neutron stars, and stellar mass black holes. Two of the leading interpretations for ULXs are X-ray binaries undergoing super-Eddington accretion onto a stellar mass compact object ($\sim10$ M$_\odot$; e.g.~Watarai et al. 2001, Stobbart et al. 2006) or an intermediate mass black hole accretor ($10^{2} \lesssim M_{\rm{BH}} \lesssim10^{5}\,M_{\odot}$;  e.g.~Colbert \& Mushotzky 1999).

Recent observations revealed the surprising discovery of ULXs that exhibit X-ray pulsations, meaning that the accretors must be rapidly rotating and magnetized neutron stars (Bachetti et al. 2014; F{\"u}rst et al. 2016; Israel et al. 2017a).  Accretion onto the neutron star in pulsar ULXs must therefore occur at super-Eddington rates. For example, P13 in NGC 7793 shows peak luminosities of $\sim100$ times the neutron star Eddington limit (Motch et al. 2014; Israel et al. 2017b). Currently, with the recent addition of NGC 300 ULX-1 (Carpano et al. 2018; Kosec et al. 2018; Walton et al. 2018; Binder et al. 2018) there are 4 bona fide ULX pulsars. The existence of these pulsar ULXs led to a dramatic paradigm shift in ULX models and underscored the need for accretion theories to explain them. Understanding the accretion and mass exchange in ULXs can provide crucial information on the evolutionary pathways that lead to binary black hole formation and gravitational wave sources. 


The leading interpretation of super-Eddington accretion models invokes a geometrically and optically thick accretion disk (e.g. Ohsuga et al. 2005; Poutanen et al. 2007; Narayan et al. 2017) as opposed to a standard optically thick but geometrically thin disc (Shakura \& Sunyaev 1973). Such geometrically thick discs with narrow funnels along the inner disc wall are believed to confine the X-ray emission to a relatively narrow opening angle and create a beaming effect where observers along the line of sight would infer greater X-ray luminosities (e.g. King 2009). However, observations of highly ionized nebulae around ULXs indicate that the surrounding gas is ionized by an isotropic X-ray radiation field consistent with observed super-Eddington luminosities (Pakull \& Mirioni 2002; Kaaret et al.~2004; Binder et al.~2018). Infrared (IR) observations of the ULX nebula surrounding Holmberg II X-1 from the \textit{Spitzer Space Telescope} (Werner et al. 2004; Gehrz et al. 2007) show high-ionization mid-IR emission lines that corroborated the interpretation of isotropic X-ray radiation field from ULXs (Berghea et al.~2010a \& b). Investigating the energetics and origins of the ULX circumstellar environment therefore provides valuable insight on the nature of supercritical accretion disks. 

Many observational studies have been conducted on optical ULX counterparts (e.g. Roberts et al. 2011: Gladstone et al. 2013). However, distinguishing the sources of optical emission from ULXs is difficult since the emission may trace the photosphere of the donor star (e.g. Motch et al. 2014), outflows from the accretion disks (e.g. Fabrika et al. 2015), or the outer accretion disk (e.g. Tao et al. 2011). To add to the complications, the outflows from ULX accretion disks exhibit emission lines in their optical spectra that resemble those of massive evolved stars like late-type nitrogen rich Wolf-Rayet stars (WNL) or luminous blue variables (LBVs) in their hot state (Fabrika et al. 2015). Additionally, most optical counterparts of ULXs are faint ($ V>24$ Vega mag) and will suffer from line-of-sight extinction due to intervening interstellar and/or local gas and dust. 

Observations at IR wavelengths, which are less affected by extinction, are less prone to causing confusion of the multiple emitting components in a ULX and therefore provide a valuable diagnostic approach to studying their mass donor stars and the circumstellar environment (e.g Heida et al 2014; Lopez et al. 2017; Lau et al. 2017). Heida et al. (2014; hereafter H14) and Lopez et al. (2017; hereafter L17) conducted systematic near-IR searches for ULX mass donors with absolute magnitudes consistent with red supergiants (RSGs), which are intrinsically bright in the near-IR. Follow-up near-IR spectroscopy by Heida et al. (2015, 2016) of three ULXs indeed revealed likely RSG mass donors. As a point of comparison, only two ULX mass donors have currently been identified from their optical counterpart (Liu et al. 2013; Motch et al. 2014). 

Nevertheless, identifying the nature of IR emission from ULXs can still be complicated due to their variability, which has been observed in the X-ray on timescales of hours to years (e.g. Heil et al. 2009). 
For example, Dudik et al. (2016) combined multi-wavelength IR to X-ray observations of the ULX Holmberg IX X-1 taken over 5 years and concluded that the origin of its mid-IR emission is either circumstellar dust or a variable jet. The uncertain origin of the mid-IR emission from Ho IX X-1, which exhibits X-ray variations by factors of 3-4 on day to year timescales (Vierdayanti et al.~2010; Walton et al. 2017), was attributed to difficulties in generating a good fit to the multi-wavelength, multi-epoch observations. 

Lau et al.~(2017) presented a \spitzer~mid-IR light curve analysis of another ULX, Holmberg II X-1, and examined the nature of its mid-IR emission. Radio observations of jet emission from Ho II X-1 performed by Cseh et al.~(2014) were serendipitously taken within a week of the mid-IR observations and allowed Lau et al.~(2017) to rule out a jet origin for the mid-IR emission in favor of circumstellar dust in the vicinity of of Ho II X-1. Radiation from the mass donor star (e.g. Heida et al. 2014) and/or circumstellar dust are therefore the most likely sources of IR emission from ULXs. 

In this paper, we present a mid-IR time domain study of \NULX~ULXs in nearby galaxies within $\sim10$ Mpc to identify mid-IR candidate counterparts and assess the nature of their IR emission. In Sec.~2 we describe our ULX sample, the multi-epoch \spitzer/IRAC imaging, ground-based near-IR observations, and follow-up, coordinated X-ray observations with the \textit{Neil Gehrels Swift Observatory}. In Sec.~3, we present the demographics of our ULX sample, descriptions of the color- and SED-based classification of stellar-like ULX IR counterpart candidates, details on the background AGN or star cluster ULX candidates, and IR non-detection limits. In Sec.~4, we discuss the transient nature of Holmberg IX X-1, the advantages of IR observations for distinguishing emitting components of ULXs, the nature of ``red" and ``blue" ULXs as seen from their IRAC color, and the implications and possible modes of dust formation from ULX systems. We summarize and conclude our results and discussion in Sec.~5. 






\section{Observations}

\subsection{Sample Selection of Ultraluminous X-ray Sources in Nearby Galaxies}

Our ULX mid-IR counterpart search consisted of \NULX~targets in the near-IR imaging campaigns conducted by \Heida~and \Lopez with the addition of the two ULXs SN~2010da/NGC 300 ULX1 and NGC 7793 P13 (Tab.~\ref{tab:ULX}). The \Heida~and \Lopez~ULXs were located in galaxies within a distance of 10 Mpc (Swartz et al. 2004, 2011; Walton et al. 2011), where RSG mass donor candidates could be identified with ground-based near-IR imaging. Building on the successful identification of near-IR counterparts by \Heida~and \Lopez, we aimed to investigate the interstellar/circumstellar medium around ULXs as well as to identify or verify their supergiant mass donor stars in the mid-IR with multi-epoch new and archival imaging observations from the \textit{Spitzer Space Telescope}.

Seven ULXs from the original \Heida~and \Lopez~sample were not included in our final ULX sample (Tab.~\ref{tab:ULX}): two were not observed by \spitzer~(Ho I XMM1 and Ho I XMM3), three had positional uncertainties greater than 7'' ([LB2005] NGC 4631 ULX1, [LB2005] NGC 891 X1, and [LB2005] NGC 891 ULX3), and the remaining two were located near the bright and saturated galaxy core in NGC 4736 (CXO J125052.7+410719 and CXO J125053.3+410714).


\begin{figure*}[t]
	\centerline{\includegraphics[width=1\linewidth]{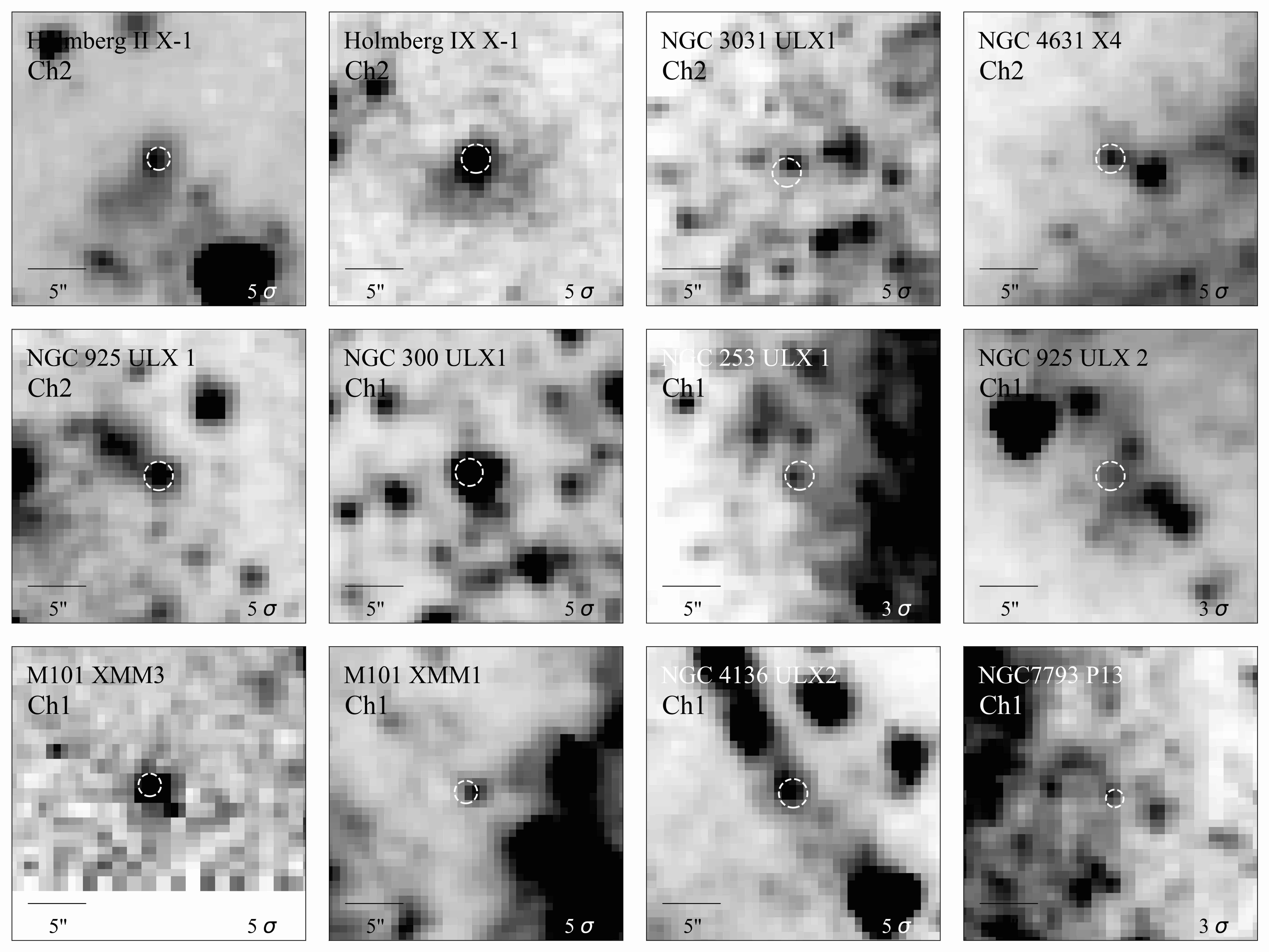}}
	\caption{\textit{Spitzer}/IRAC $24 \times 24''$ cutouts of the \NIRULX~stellar ULX mid-IR candidate counterparts overlaid with a white, dashed circle centered on the X-ray derived position of the ULX. The radius of the white dashed circles indicates the $2\sigma$ positional uncertainty of the ULX, which includes both X-ray and mid-IR \spitzer~positional uncertainties. The $\sigma$ value on the bottom right corner of each cutout indicates the value above the median flux adopted for the upper limit of the brightness stretch. The orientation of each cutout is aligned with the detector coordinates at the time of the observation.}
	\label{fig:ULX_dets}
\end{figure*}

\begin{figure*}[!t]
	\centerline{\includegraphics[width=1.0\linewidth]{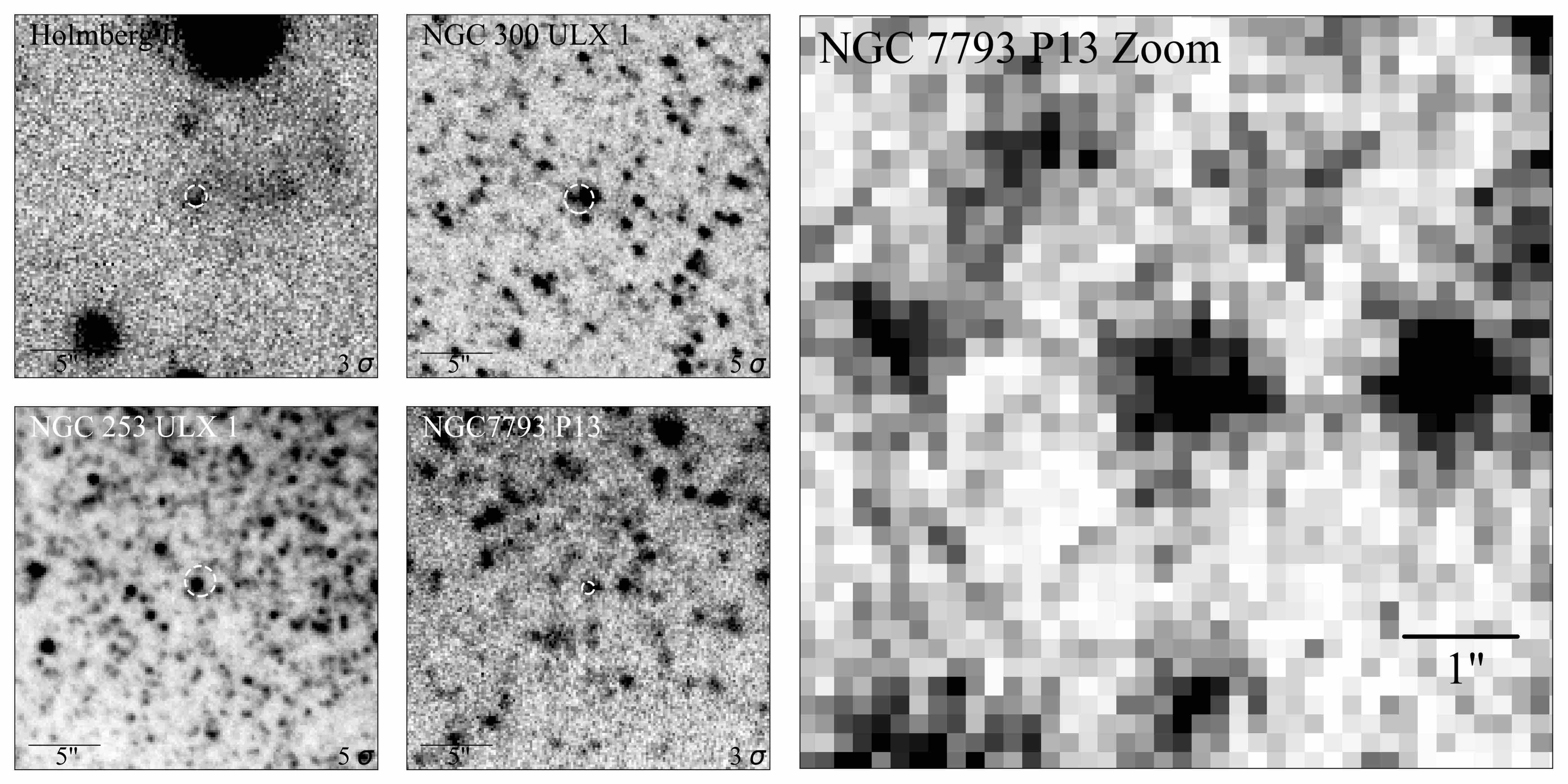}}
	\caption{(\textit{left}) Ground-based $K_s$-band imaging observations of (\textit{top left}) Holmberg II X-1 taken by Keck/MOSFIRE, (\textit{top right}) NGC 300 ULX1 taken by Magellan/FourStar, (\textit{bottom left}) NGC 253 ULX1 taken by Magellan/FourStar, and (\textit{bottom right}) NGC 7793 P13 taken by Magellan/FourStar. The dashed circles correspond to the 2$\sigma$ positional uncertainties of the ULX's X-ray coordinates. (\textit{right}) Zoomed $K_s$-band Magellan/FourStar image of NGC 7793 centered on P13. North is up and east is to the left. }
	\label{fig:ULX_NIR}
\end{figure*}

\subsection{Mid-Infrared Observations with \textit{Spitzer}}
\subsubsection{SPIRITS}

In order to search for mid-IR ULX counterparts, we utilized imaging data taken in the SPitzer InfraRed Intensive Transient Survey (SPIRITS; \citealt{Kas2017}) as well as archival \textit{Spitzer} Post Basic Calibrated (PBCD) imaging data from the \textit{Spitzer Heritage Archive}\footnote{\url{https://sha.ipac.caltech.edu/applications/Spitzer/SHA/}}. SPIRITS targeted 194 nearby galaxies within 20 Mpc with observations performed in the 3.6 $\mu$m and 4.5 $\mu$m bands of the InfraRed Array Camera (IRAC, Fazio et al. 2004) on-board warm \textit{Spitzer} with cadence baselines ranging from 1 week to 6 months. Each SPIRITS observation is a mosaic of seven dithered 100 s exposures in each IRAC filter and achieves a $5\sigma$ depth of 20 mag at [3.6] and 19.1 mag at [4.5] on the Vega system.

SPIRITS is an on-going program since 2014 and has recently been renewed for \spitzer~Cycle 14, which runs until 30 November, 2019. Additionally, many of the nearby ULX host galaxies have previous imaging observations with \spitzer~from as long ago as 2004. For some targets, we therefore have $>10$ yr baselines. We note that the imaging depth from archival \spitzer~data taken outside of SPIRITS varies with each program given their different science goals for imaging identical fields of our target galaxies. The depth and cadence of SPIRITS enables multi-epoch $5\sigma$ detections of mid-IR ULX counterparts with absolute magnitudes brighter than $M_{[3.6]} \approx -10$ and $M_{[4.5]}\approx-11$ in galaxies $\lesssim 10$ Mpc away. This detection limit is consistent with the observed mid-IR \spitzer~magnitudes of supergiants such as RSGs, LBVs, and sgB[e]s (\citealt{B2009}). 


\begin{turnpage}
\begin{figure*}[t]
	\centerline{\includegraphics[width=1\linewidth]{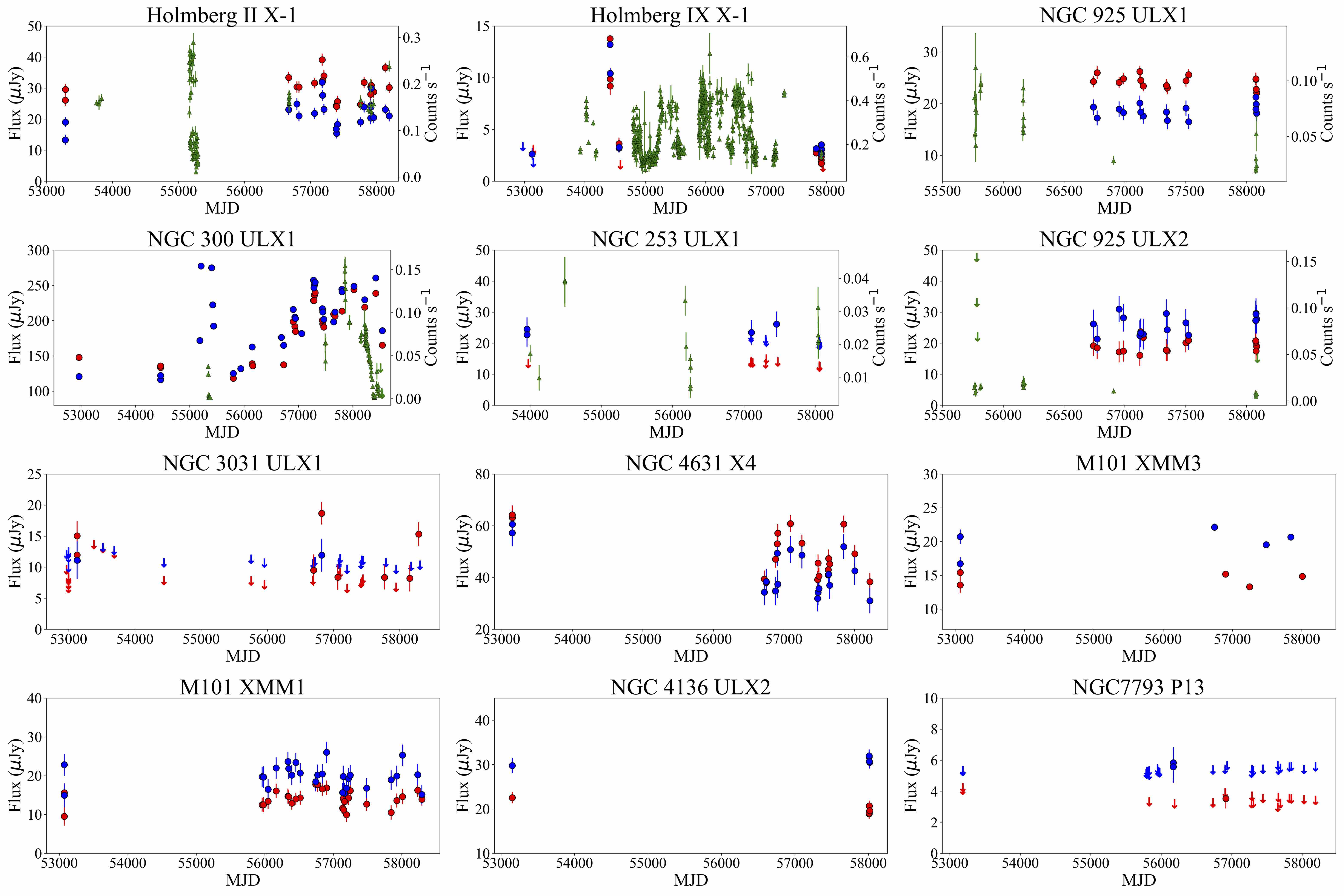}}
	\caption{\spitzer/IRAC light curves of the \NIRULX~stellar ULX mid-IR candidate counterparts taken from SPIRITS and archival images on the \textit{Spitzer Heritage Archive}. The blue and red points correspond to the Ch1 (3.6 $\mu$m) and Ch2 (4.5 $\mu$m) flux, respectively, in units of $\mu$Jy. Swift/XRT light curves (green triangles) are plotted for 6 of the ULX mid-IR candidates. Coordinated IRAC and XRT observations of Holmberg II X-1, Holmberg IX X-1, NGC 925 ULX1, NGC 925 ULX2, and NGC 253 ULX1 were performed around MJD $\sim58000$. The error bars on the points correspond to the $1\sigma$ flux uncertainties. $3\sigma$ upper limits for IRAC and $1\sigma$ upper limits for XRT are indicated by the arrows for non-detections.}
	\label{fig:ULX_LCs}
\end{figure*}
\end{turnpage}

\subsection{Searching for Short-Term Variability}

We searched for short-term mid-IR variability of 6 ULXs with \spitzer/IRAC (PID - 13163) at the same depth as SPIRITS with cadence baselines ranging from 0.5 days to 1 week. In this 10 hr \spitzer~Director's Discretionary Time program, we performed 4 imaging observations of each ULX with IRAC Channels 1 and 2 taken at intervals of 0.5 days, 1 day, and $\sim1$ week at the same depth of the SPIRITS imaging. The following ULXs were observed in this program: Holmberg II X-1, Holmberg IX X-1, NGC 253 ULX1, NGC 4136 ULX2, NGC 925 ULX1, and NGC 925 ULX2.

\subsubsection{Searching for Mid-IR Counterparts}
We searched for 3.6 and 4.5 $\mu$m mid-IR counterpart candidates with a signal-to-noise ratio $>3$ at the X-ray positions of the \Heida~ and \Lopez~ULXs within an error radius determined from the $2\sigma$ error circle from the X-ray positions and twice the maximum RMS uncertainty quoted by the IRAC handbook, $\sigma_\mathrm{RMS} \lesssim 0.3''$. Multi epoch \spitzer~images that covered overlapping fields of view exhibited variations in the astrometry by up to $\sim 0.6''$, which is on the approximate size of a PBCD mosaic pixel. Such offsets are in line with the positional uncertainties indicated in the IRAC instrument handbook ($\sigma_\mathrm{RMS} \lesssim 0.3''$). We therefore adopted an astrometric uncertainty of $0.6''$ for the \spitzer/IRAC images. We constrained our search to Channel 1 (3.6 $\mu$m) and Channel 2 (4.5 $\mu$m) IRAC images since these wavelengths provide the deepest sensitivity limits to detect emission from supergiant mass donors with \spitzer. Channels 1 and 2 also have the longest temporal baseline coverage due to SPIRITS, which allows us to investigate the variability from ULX mid-IR counterparts. 

We searched for point source-like mid-IR counterparts from IRAC imaging observations taken closest in time to the near-IR observations by H14 and L17. For mid-IR non-detections of the ULXs located within the SPIRITS galaxies, we scanned for transient or variable behavior in the multi-epoch imaging taken in SPIRITS. Aperture photometry was performed on sources with positive mid-IR detections within the error circle. An aperture radius of 4.0 PBCD mosaic pixels (2.4'') and a sky background annulus extending from 4 to 12 mosaic pixels (2.4'' to 7.2'') were used to measure the mid-IR flux. Given the larger extent of the \spitzer/IRAC pixel response functions (PRF), we adopted aperture correction factors from the \textit{Spitzer}/IRAC manual of 1.215 and 1.233 for Channels 1 and 2, respectively. For the three sources NGC 4631 X4, NGC 253 ULX1 and NGC 7793 P13, we used a smaller aperture of 2.3 mosaic pixels (1.4'') and aperture corrections of 1.71 and 1.76 for Channels 1 and 2 (Timlin et al.~2016), respectively, due to crowding by nearby stars and extended background structures.

\subsection{Near-Infrared Ground-Based Observations}

As a part of SPIRITS (Kasliwal et al.~2017), we performed follow-up ground-based near-IR \textit{J, H,} and/or $K_s$ imaging observations with the Multi-object Spectrometer for Infra-red Exploration (MOSFIRE; McLean et al. 2010, 2012) on Keck I, FourStar (Persson et al. 2013) on the Magellan 6.5 Baade Telescope, and the Wide Field Infrared Camera (WIRC; Wilson et al. 2003) on the Palomar 200-inch Telescope. Using these observations we measured the near-IR photometry for the mid-IR ULX counterparts with fluxes consistent with supergiants. These observations also provide additional imaging epochs that can be compared against the photometry obtained by H14 and L17. In Table~\ref{tab:ULXNIRTab}, we list the near-IR photometry of the ULX supergiant candidates in the following galaxies: Holmberg II, M81, NGC 253, NGC 300, NGC 4631, and NGC 7793.

In a dedicated IR ULX follow-up program, the ULX Holmberg II X-1 was observed with the NIRC2 instrument on Keck II with the laser guide star adaptive optics (LGS AO) system on 2018 Nov 28. Holmberg II X-1 was imaged with the wide field camera ($40'' \times 40''$, $0.04''$ pix$^{-1}$) in $H$ and $K_s$ bands in a 3-point dither pattern with 10'' offsets. At both $H$ and $K_s$ bands, 30 second integration times were used per exposure. The total integration times at $H$ and $K_s$ bands were 1800 and 1440 sec, respectively. A bright ($R\sim13.3$ mag) source located 34'' from Holmberg II X-1 was used as a Tip-Tilt star in LGS-mode, which allowed us to achieve an angular resolution of $\sim0.16$''. 

Dark and dome flat frames were obtained at the end of the night and used to produce dark-subtracted and flat-fielded images in $H$ and $K_s$ bands. Since Holmberg II X-1 was not detected in the single 30 second exposures, the centroid of a bright source within the 40'' field of view was used to align the individual exposures before coadding. A near-IR photometric standard star, P035-R (Persson et al.~1998), was observed in the middle of the night to perform the photometric calibration at $H$ and $K_s$ bands. The zero-point magnitudes were 25.33 and 24.46 at $H$ and $K_s$ bands, respectively. In order to measure the emission from the unresolved Holmberg II X-1 point source within its extended nebulosity, a small $0.2''$ radius aperture was used with aperture correction factors of $3.47 \pm0.27$ and $2.18\pm0.2$ assuming all of the emission is contained with a $1''$ radius for the $H$ and $K_s$ band images, respectively. These aperture corrections factors were derived by measuring the mean of the flux ratios of 4 sources of similar brightness to Holmberg II X-1 with $1''$ vs $0.2''$ photometric aperture radii.

\subsection{Neil Gehrels Swift Observatory Follow up}

In order to investigate possible correlations between mid-IR/X-ray variability in ULXs from our \spitzer~ DDT program (PID - 13163), we conducted contemporaneous observations with the X-ray Telescope (XRT) on the Neil Gehrels Swift Observatory (Gehrels et al.~2004) of 5 ULXs: Holmberg II X-1, Holmberg IX X-1, NGC 925 ULX1, NGC 925 ULX2, and NGC 253 ULX1. We also searched for long-term $\sim$yr-long correlations from the ULX Holmberg II X-1 with near-contemporaneous Swift/XRT and \spitzer~observations in SPIRITS. Although the timing of previous Swift/XRT observations of these five ULXs did not always overlap with \spitzer/IRAC observations, we included all previous XRT observations for our analysis. We also utilized the available XRT observations of SN~2010da/NGC 300 ULX1 to compare with the mid-IR observations from SPIRITS. 






\begin{figure*}[t]
	\centerline{\includegraphics[width=1\linewidth]{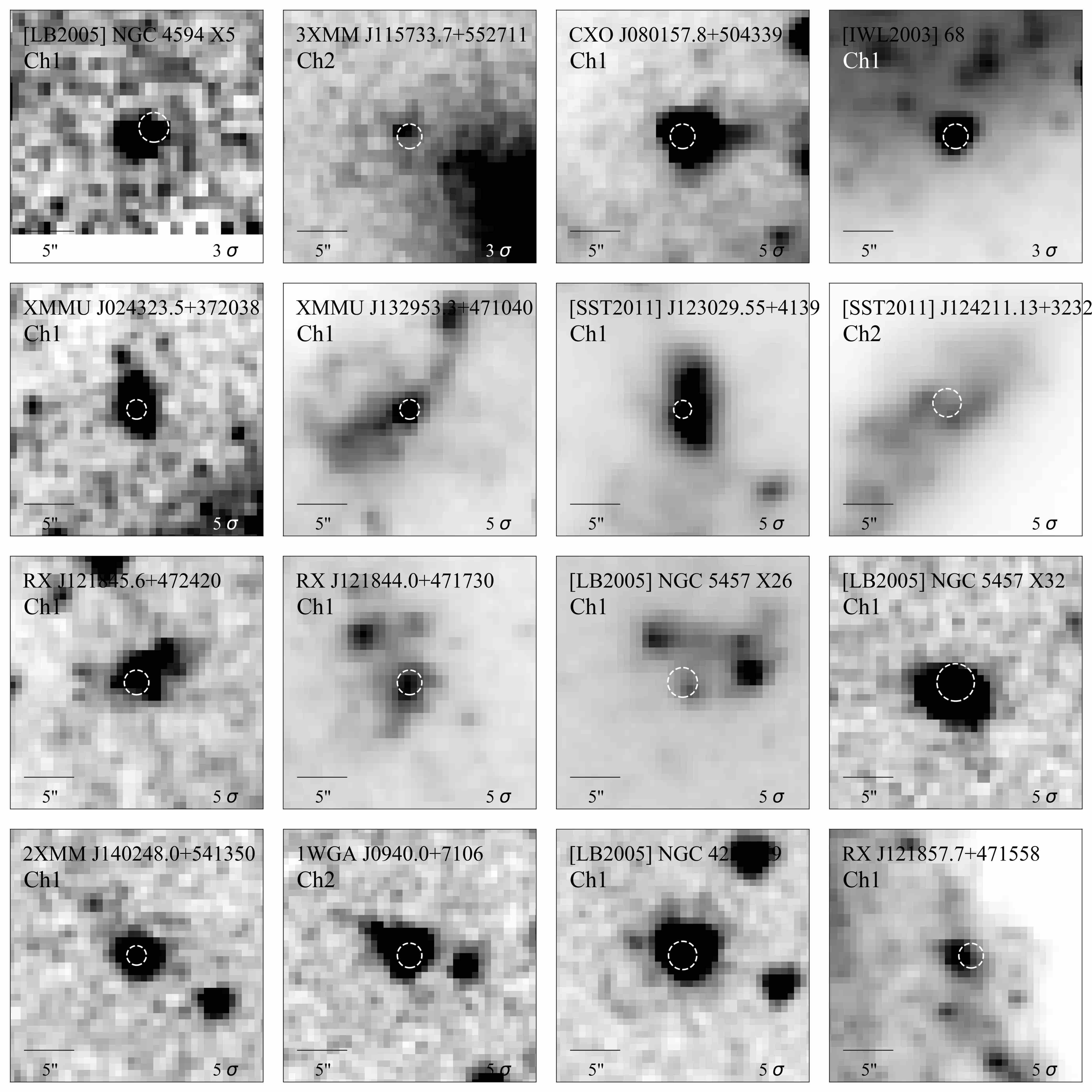}}
	\caption{\textit{Spitzer}/IRAC $24 \times 24''$ cutouts of the \ODULX~ULX mid-IR candidate counterparts that are likely star clusters, AGN, or background galaxies overlaid with a white, dashed circle centered on the X-ray derived position of the ULX. The radius of the white dashed circles indicates the $2\sigma$ positional uncertainty of the ULX, which includes both X-ray and mid-IR \spitzer~positional uncertainties. The orientation of each cutout is aligned with the detector coordinates at the time of the observation.}
	\label{fig:ULX_ODets}
\end{figure*}

\section{Results and Analysis}

\subsection{Demographics}
In our search of \NULX~ULXs we find \NIRULX~stellar mid-IR counterpart candidates consistent with the absolute magnitudes of supergiants, [3.6] $> -12.5$ and [4.5] $> -13.0$. 5 of these \NIRULX~ULXs exhibited one or more epochs of significant ($>3\sigma$) variability at both 3.6 and 4.5 $\mu$m. All but three of the mid-IR counterparts have been previously detected in the near-IR. These three mid-IR sources with no detected near-IR counterparts, Holmberg IX X-1, NGC 3031 ULX1, and NGC 4631 X4, exhibit variability in the mid-IR. 

Cutouts of the highest significance detections of the \NIRULX~stellar mid-IR ULX counterparts are shown in Fig.~\ref{fig:ULX_dets} and are overlaid with the error circle of the ULX based on its X-ray position and the \spitzer/IRAC astrometric uncertainty. The detected near-IR counterparts are shown in Fig.~\ref{fig:ULX_NIR}. The \spitzer~3.6 and 4.5 $\mu$m light curves of the \NIRULX~counterparts are shown in Fig.~\ref{fig:ULX_LCs} and roughly span between 2004 and 2018. Figure ~\ref{fig:ULX_LCs} also shows X-ray light curves measured by \textit{Swift}/XRT of Holmberg II X-1, Holmberg IX X-1, NGC 300 ULX1, NGC 925 ULX1, NGC 925 ULX2, and NGC 253 ULX1 overlaid on the mid-IR light curves. The stellar mid-IR ULX counterpart candidates are discussed in detail in Sec.~\ref{sec:ULX_SC}.

In addition to the \NIRULX~ stellar candidates, \ODULX~ULXs have mid-IR counterpart candidates that exhibit absolute magnitudes brighter than that expected from supergiants, [3.6] $< -12.5$ and [4.5] $< -13.0$ (Fig.~\ref{fig:ULX_ODets}). These bright mid-IR counterparts are likely associated with star clusters or background active galactic nuclei (AGN) and discussed in detail in Sec.~\ref{sec:ULX_AGN}.

A majority of the ULXs in our sample, \NDULX, do not have a mid-IR detection with \spitzer/IRAC. For these non-detections, we provide $3\sigma$ upper limits at the \spitzer~epochs closest in time to the respective near-IR imaging observations performed by \Heida~and \Lopez and are provided in Tab~\ref{tab:ULXNonDetTab}. The average absolute magnitude $3\sigma$ non-detection limits were $-11.6\pm1.9$ and $-11.8\pm1.8$ for [3.6] and [4.5], respectively. These limits are slightly brighter than the estimated depth of SPIRITS due to the crowding and high background emission at the location of most of the ULXs.

\begin{figure}[t]
	\centerline{\includegraphics[width = 1\linewidth]{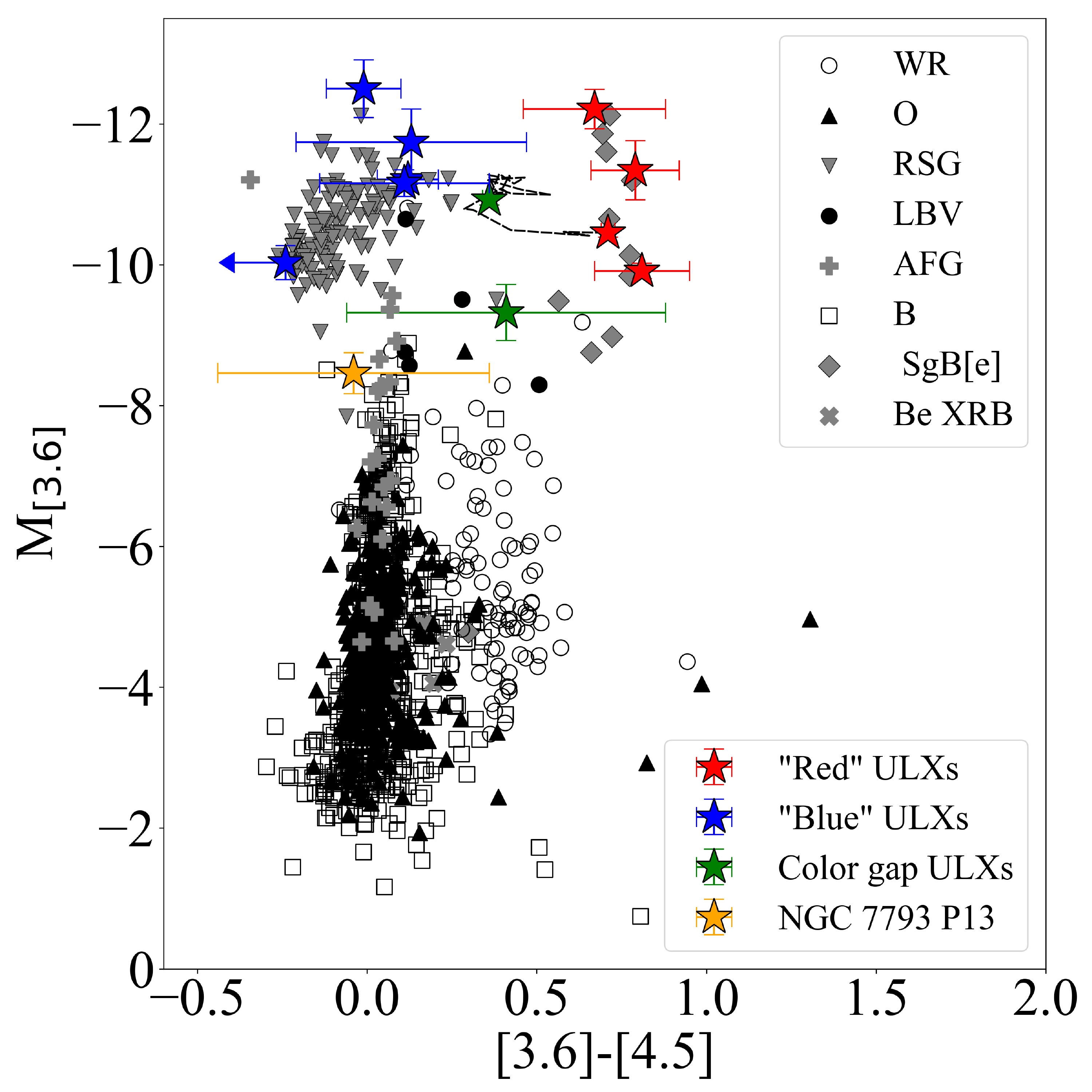}}
	\caption{\spitzer/IRAC color-magnitude diagram of 11 of the 12 stellar mid-IR ULX counterpart candidates (star symbols) plotted over massive stars in the LMC of known spectral type compiled by Bonanos et al.~(2009). The progenitor (MJD 52964) and the most recent observation of NGC 300 ULX-1 (MJD 57802) are plotted and connected by a dashed line showing its color evolution from redder to bluer. Four ULX counterparts exhibited ``red" colors  of $[3.6]-[4.5] \sim 0.8$: NGC 300 ULX1 (MJD 52964), Holmberg II X-1, NGC 925 ULX1, and NGC 4631 X4. Seven counterparts exhibited ``blue" colors of $[3.6]-[4.5] \sim 0$: NGC 253 ULX1, NGC 925 ULX2, M101 XMM1, M101 XMM3, NGC 4136 ULX2, and NGC 7793 P13. The two ULX counterparts that are not consistent with these colors are NGC 300 ULX1 (MJD 57802) and NGC 3031 ULX1. Holmberg IX X-1 is not shown on this plot due to its transient nature.}
	\label{fig:ULX_CMD}
\end{figure}

\begin{figure*}[t!]
	\centerline{\includegraphics[width=1\linewidth]{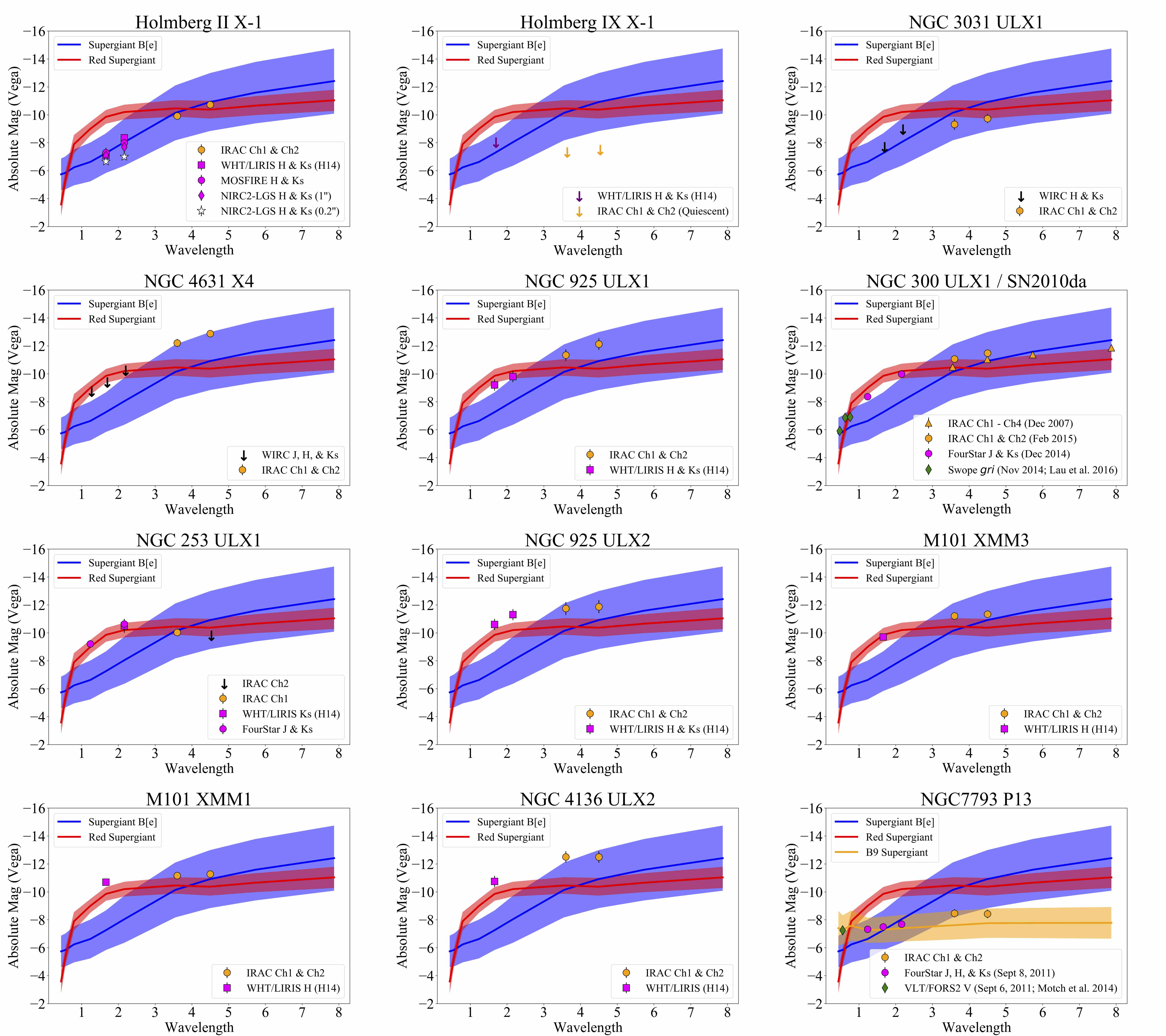}}
	\caption{Infrared Spectral Energy Distributions (SEDs) of the 12 stellar mid-IR ULX counterpart candidates plotted against SED templates of 122 RSG and 11 sgB[e] stars in the LMC cataloged by Bonanos et al.~(2009). The wavelength on the X-axis is shown in units of $\mu$m. The template of a 18 B9 supergiant is also plotted in the SED of NGC 7793 P13, whose mass donor star was identified as a B9 supergiant by Motch et al.~(2014). The solid lines correspond to the median VIJHKs and IRAC of the SED template stars and the surrounding shaded regions indicate the 1$\sigma$ spread in the magnitudes of the distribution.}
	\label{fig:ULX_SEDs}
\end{figure*}

\subsection{Mid-IR Color Magnitude Diagram}

The catalogue of mid-IR stellar properties compiled by Bonanos et al.~(2009) provides a valuable resource to identify the stellar types of ULX donor stars from their mid-IR colors and magnitudes (e.g. Williams \& Bonanos 2016). This catalogue consists of uniform photometry from 0.3 to 24 um of 1750 massive stars of known spectral type in the Large Magellanic Cloud (LMC), with the mid-IR measurements taken from the Spitzer program "Surveying the Agents of a Galaxy's Evolution (SAGE; Meixner et al.~2006). Figure~\ref{fig:ULX_CMD} shows the stellar ULX mid-IR candidates on a \spitzer/IRAC mid-IR color-magnitude diagram (CMD) of spectrally classified massive stars in the LMC from Bonanos et al. (2009). 

The [3.6] - [4.5] colors of the stellar mid-IR ULX counterparts brighter than $M_\mathrm{3.6}<-9$ exhibit an apparent bi-modal distribution, with 4 that are ``red'' ([3.6] - [4.5] $\sim 0.8$) and 6 that are ``blue'' ([3.6] - [4.5] $\sim 0$). The groupings of the red and blue ULXs are consistent with the CMD distribution of supergiant B[e] (sgB[e]) stars and red supergiants (RSGs), respectively. Notably, three of these ULXs that are consistent with RSGs in the CMD--NGC 253 ULX1, NGC 925 ULX2, and NGC 4136 ULX2--were previously identified spectroscopically as RSGs based on CO absorption features in their H-band continuum (Heida et al.~2015, 2016). The ``red" [3.6] - [4.5] color from the sgB[e] stars arises from hot circumstellar dust at $\sim600$ K (Bonanos et al.~2009), whereas the ``blue" color from the RSGs is likely due to molecular absorption features around 4.5 $\mu$m on the photospheric continuum emission (e.g. Verhoelst et al. ~2009). 

In Fig.~\ref{fig:ULX_CMD}, the first and most recent \spitzer/IRAC observations of NGC 300 ULX1 are shown to demonstrate the significant change in its mid-IR color and magnitude. Its color before the SN-impostor explosion in 2010 appeared sgB[e]-like, while the most recent observation of NGC 300 ULX1 taken on 2018 November 3 show that its color falls in between the RSGs and sgB[e]s. In Sec.~\ref{sec:ULX_SC}, we describe the mid-IR color and magnitudes of the individual candidate stellar ULX counterparts in further detail.

\subsection{Spectral Energy Distributions}

Photometry of the spectrally classified massive stars in the LMC from Bonanos et al.~(2009) provides IR spectral energy distribution (SED) ``templates" to compare to the stellar mid-IR ULX candidates. RSG, sgB[e], and late B (8 and 9) supergiant SED templates are created using the median \textit{VIJH}K$_s$+IRAC photometry of the known populations in the LMC. 122, 11, and 18 stars were used to construct the RSG, sgB[e], and late B supergiant SED templates, respectively. While the mid-IR colors of RSGs and sgB[e]s are more distinct in the CMD (Fig.~\ref{fig:ULX_CMD}), the SED templates help to distinguish the spectral shape of the stellar candidates in the IR. In addition to the mid-IR, near-IR photometry from H14, L17 and this work and optical photometry from Motch et al.~(2014) and Lau et al.~(2016) are plotted in the IR SEDs to compare against the templates (Fig.~\ref{fig:ULX_SEDs}). Notably, the IR SEDs of Holmberg II X-1 and NGC 253 ULX1 show a close agreement to their previous classifications as a sgB[e]-like (Lau et al. 2017b) and RSG-like (Heida et al. 2015), respectively.

\subsection{ULX Stellar Candidates}
\label{sec:ULX_SC}

\subsubsection{SN2010da/NGC 300 ULX1}

NGC 300 ULX1 is one of four known pulsar ULXs and was previously identified as the impostor supernova (SN) 2010da (Monard 2010, Chornock \& Berger 2010; Elias-Rosa et al. 2010a,b). SN2010da ``exploded' in 2010 May 24 in the nearby spiral galaxy NGC 300 (d = 1.88 $\pm0.05$ Mpc; Gieren et al. 2005); however, its optical, IR, and X-ray emission persisted and is still detected in the most recent observations (Binder et al.~2011, 2016, 2018; Villar et al. 2016; Lau et al. 2016). Binder et al.~(2016) present multi-epoch \textit{XMM-Newton}, \textit{Swift}, and \textit{Chandra} observations of SN2010da that revealed an increasing X-ray luminosity despite no detected X-ray emission before the 2010 outburst. The non-detection of X-ray emission previous to the 2010 outburst suggests the transient could be linked to the formation of an X-ray binary (Binder et al.~2016). Interestingly, IR-luminous outbursts have been suggested as tracers of common envelope events in the formation of high-mass X-ray binaries (Oskinova et al.~2018).

Initially, the X-ray behavior from SN2010da was interpreted by Binder et al.~(2016) as recurring outburst activity from a young ($\lesssim 5$ Myr old) HMXB with a luminous blue variable (LBV) stellar companion. Based on optical and IR observations, Lau et al. (2016) suggest the stellar counterpart is a sgB[e] star but did not rule out the LBV hypothesis. Villar et al. (2016) argue the counterpart is a yellow supergiant. Both Lau et al. (2016) and Villar et al. (2016) agree that the progenitor was enshrouded by dust that was mostly destroyed in the 2010 outburst. 

Recent \textit{Chandra} and \textit{NuSTAR} observations discovered that SN2010da hosts a pulsing neutron star that exhibits super-Eddington X-ray luminosities of a few $\times10^{39}$ ergs s$^{-1}$ (Carpano et al. 2018, Walton et al. 2018, Kosec et al. 2018). After its discovery as a ULX, SN2010da was re-designated as NGC 300 ULX1. Follow-up optical spectroscopy with Gemini GMOS spectroscopy of NGC 300 ULX1 by Binder et al. (2018) show from the X-ray ionized HeII$\lambda4686$ emission line that geometric beaming effects are minimal and that NGC 300 ULX1 is a bona fide ULX. 

Continued monitoring of NGC 300 ULX1 with \spitzer/IRAC beyond the light curves presented by Lau et al. (2016) and Villar et al. (2016) show that since the 2010 outburst its mid-IR counterpart has gradually brightened to a factor of $\sim2$ brighter than its pre-2010 appearance but faded significantly in the most recent observation taken in 2019 March (MJD 58546) (Fig.~\ref{fig:ULX_LCs}). The median absolute magnitudes exhibited by NGC 300 ULX1 post-2010 outburst are $[3.6]\sim-11.1$ and $[4.5]\sim-11.5$, which are consistent with mid-IR brightness of RSGs and sgB[e]s. Before the 2010 outburst, the mid-IR color and magnitude of NGC 300 ULX1 were consistent with that of sgB[e]s. As shown in Fig.~\ref{fig:ULX_CMD}, after the 2010 outburst NGC 300 ULX1 exhibited a ``blueward" shift in its mid-IR color into the color gap between sgB[e] stars and RSGs. 

Based on the SED templates (Fig.~\ref{fig:ULX_SEDs}), 3.6 - 8.0 $\mu$m photometry taken before the 2010 outburst by Cold \spitzer/IRAC show a closer agreement with the sgB[e] template than that of an RSG. However, post-outburst near-IR photometry of NGC 300 ULX1 from late 2014/early 2015 show an IR SED that appears more consistent with the RSG template. Post-outburst optical photometry from the Swope 1-m Telescope at Las Campanas Observatory taken in late 2014 (Lau et al.~2016) show a $g$-band excess relative to the RSG template but a consistent match to the sgB[e] template. The $r$- and $i$-band photometry are consistent within uncertainties to both RSG and sgB[e] templates, but we note that the $r$-band photometry includes the strong H$\alpha$ emission observed from NGC 300 ULX1 (e.g. Binder et al.~2018). The RSG-like near-IR photometry and the blueward shift in its mid-IR colors after the 2010 outburst suggests NGC 300 ULX1 may host a RSG mass donor, where the $g$-band excess may be due to the accretion disk (See Sec.~\ref{Sec:DDD}).

\begin{figure}[t]
	\centerline{\includegraphics[width=1.0\linewidth]{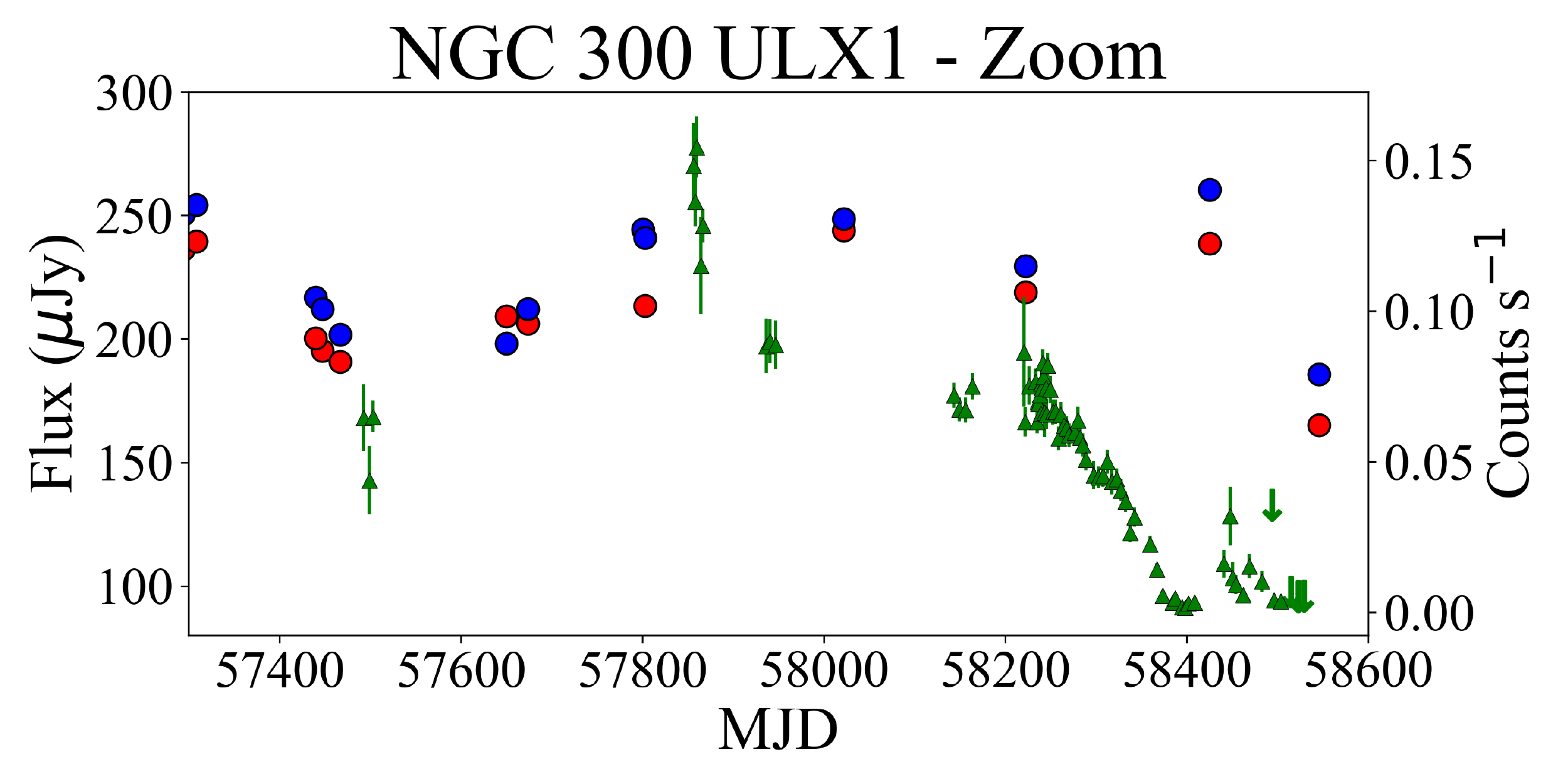}}
	\caption{Zoomed-in mid-IR and X-ray light curve of NGC 300 ULX1 taken between MJD 57300 - MJD 58550. The blue and red circles correspond to the Ch1 and Ch2 fluxes measured by \spitzer/IRAC, and the green triangles represent the X-ray flux measured by Swift/XRT.}
	\label{fig:NGC300_zoom}
\end{figure}

\spitzer~and \textit{Swift} observations of NGC 300 ULX1 taken between 2016 and mid 2018 (MJD $\sim57500-58200$; Fig.~\ref{fig:NGC300_zoom}) show that the mid-IR and X-ray emission may be correlated. Both mid-IR and X-rays show a peak in emission at around 2017 April (MJD $\sim57850$), where the mid-IR flux increased by $\sim20\%$ while the X-ray counts increased by up to a factor of $\sim2$ to $0.136\pm0.014$ counts s$^{-1}$. During this X-ray peak, the mid-IR color also became bluer by $\Delta ([3.6] - [4.5]) = 0.05\pm0.01$ from the mid-IR observations taken around January 2016 (MJD$\sim57400$). \textit{Swift/XRT} observations taken around late 2018 March (MJD$\sim58200$) show a factor of $\sim2$ decrease in count rate from the 2017 April peak. This was followed by a decline over the following 200 days where the lowest observed count rate was $0.0017\pm0.0009$ on 2018 October 7 (MJD 58398), almost an order of magnitude lower from the X-ray peak. 

No \spitzer~images of NGC 300 ULX1 were taken during the 200 day decline in the X-ray count rate, but observations taken on 2018 November 3 (MJD 58425) show that the mid-IR emission increased relative to epochs before the onset of the fading X-ray emission. \textit{Swift/XRT} observations taken 15 (MJD 58440) and 22 (MJD 58447) days after the 2018 November 3 \spitzer~imaging revealed a small X-ray peak, which was followed by small-amplitude variability and non-detections. During this small X-ray peak on MJD 58447, the X-ray emission increased by over an order of magnitude ($0.032\pm0.012$ counts s$^{-1}$) from the X-ray minimum measured on MJD 58398. The most recent \spitzer~observation from 2019 March 3 (MJD 58546) showed a decrease in the mid-IR flux by $\sim40\%$ from the preceding 2018 November 3 measurement. With the current data, it is unclear whether or not the mid-IR and X-ray emission from NGC 300 ULX1 are correlated but continued \spitzer~monitoring is planned throughout \spitzer~Cycle 14.




\begin{figure*}[t!]
	\centerline{\includegraphics[width = 1\linewidth]{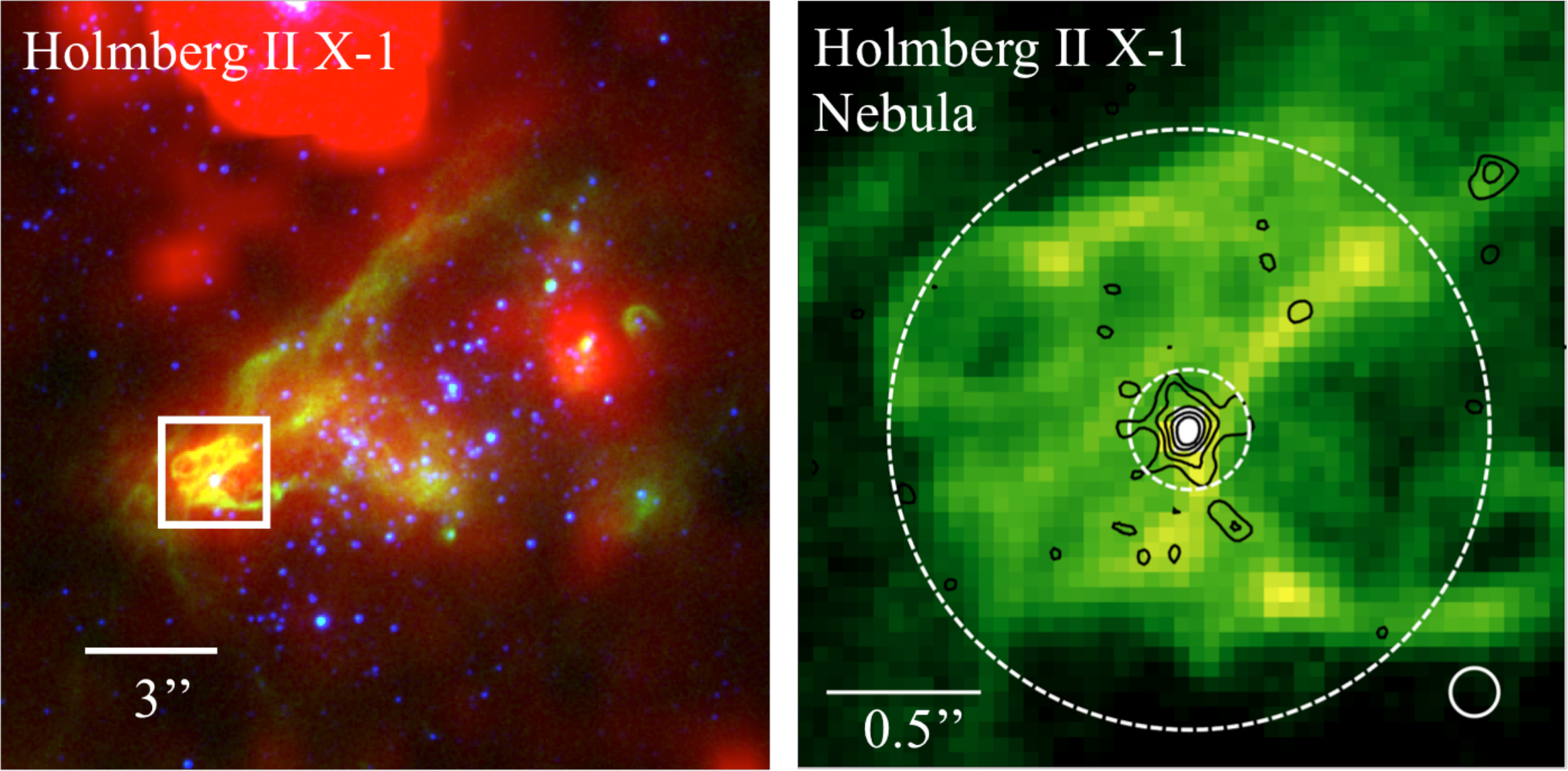}}
	\caption{(\textit{left}) False color \spitzer/IRAC 4.5 $\mu$m (Ch2, red) and HST/WFC H$\alpha$ (F658N, green) and V-band (F550M, blue) image of Ho II X-1 and its surrounding 17'' ($\sim300$ pc) environment. The overlaid white box is centered on Ho II X-1 and its $\sim1''$ ($\sim20$ pc) nebula. (\textit{right}) Zoomed image of the region in the white box in (\textit{right}) showing the HST/WFC H$\alpha$ emission (green) overlaid with NIRC2-LGSAO K$_s$-band contours corresponding to 5$\sigma$, 7$\sigma$, 9$\sigma$, 11$\sigma$, and 13$\sigma$. The small and large dashed, white circles correspond to the $0.2''$- and $1.0''$-radius apertures used to measure the near-IR NIRC2 emission from the central ULX and the extended nebula, respectively. The approximate beam size of the NIRC2 K$_s$-band image ($\sim0.16''$) is shown at the bottom right. North is up and east is to the left.}
	\label{fig:HoII_FC}
\end{figure*}

\subsubsection{Holmberg II X-1}


Holmberg II X-1, hereafter referred to as Ho II X-1, is a well-known ULX that exhibits X-ray ($\sim$0.3 - 10 keV) luminosities in excess of $10^{40}$ erg s$^{-1}$ and is located in the dwarf galaxy Holmberg II (d = 3.39 Mpc; e.g.~Pakull \& Mirioni 2002; Karachentsev et al.~2002). Ho II X-1 is surrounded by a $\sim10$ pc-sized X-ray ionized nebula located in the ``heel'' of the $\sim100$ pc Foot Nebula (Fig.~\ref{fig:HoII_FC}; Pakull \& Mirioni 2002; Kaaret et al. 2004). Multi-epoch radio observations of Ho II X-1 revealed variability and spatially extended non-thermal emission consistent with a jet ejection (Cseh et al.~2014, 2015). H14 initially classified the IR counterpart of Ho II X-1 a RSG donor star based on its $H$ and $K_s$-band absolute magnitude. Follow-up $H$-band spectroscopy by Heida et al.~(2016) did not show any absorption features indicative of an RSG photosphere, but instead revealed [Fe II] ($\lambda$1.644) and hydrogen Brackett emission lines. Lau et al.~(2017b) presented mid-IR photometry from \spitzer~that showed its mid-IR colors and magnitude were consistent with that of an sgB[e] star. Although the non-thermal jet seen in the radio observations may be expected to contribute to the mid-IR emission, Spitzer observations serendipitously taken within a week of one of the radio observations showed that the mid-IR brightness and spectral shape of Ho II X-1 were inconsistent with a jet origin. (Lau et al.~2017b).


Deep near-IR $H$ and $K_s$-band imaging observations of Ho II X-1 from Keck/MOSFIRE show consistent photometry with H14 (See Tab.~\ref{tab:ULXNIRTab}). High angular resolution H and $K_s$-band imaging of Ho II X-1 with Keck/NIRC2-LGSAO (Fig.~\ref{fig:HoII_FC}, \textit{right}) show similarly consistent H-band photometry with a $1''$-radius aperture  ($H=20.42\pm0.06$) but slightly fainter K-band emission ($K_s=19.81\pm0.05$; See Fig.~\ref{fig:ULX_SEDs}). The $H$ and $K_s$-band emission of the central $0.2''$ around Ho II X-1 account for $67\pm5\%$ and $53\pm5\%$ of the emission within a $1''$ radius, respectively. The IR SED of Ho II X-1 (Fig.~\ref{fig:ULX_SEDs}) shows a consistency with the sgB[e] template and strengthens the interpretation of the mid-IR counterpart as a sgB[e]-like star or, more broadly, that it is a ULX exhibiting the ``B[e] phenomenon" (See Sec.~\ref{sec:ULXBe}).



Lau et al. (2017b) also revealed the variable emission from the mid-IR counterpart of Ho II X-1. In this work, the continued \spitzer/IRAC monitoring of Ho II X-1 shows on-going fluctuations in the emission from its mid-IR counterpart (see Fig.~\ref{fig:ULX_LCs}). The median 3.6 and 4.5 $\mu$m flux measured from Ho II X-1 is $22.5\pm2.3$ $\mu$Jy ($17.74\pm0.11$ mag) and $30.4\pm2.3$ $\mu$Jy ($16.93\pm0.08$ mag), respectively. Observations taken 2018 Jan 15 (MJD 58133.4) show a $>3\sigma$ enhancement in the 4.5 $\mu$m emission above the median flux, similar to the peak 4.5 $\mu$m flux measured on 2015 Jun 4 (MJD 57177.29). However, the contemporaneous 3.6 $\mu$m flux only increased by $\sim1\sigma$ from the median during the 2018 Jan 15 flux peak, whereas the 3.6 $\mu$m flux increased by $\sim5\sigma$ above the median during the 2015 Jun 4 peak. 

As first shown by Grise et al.~2010, \textit{Swift}/XRT monitoring of Ho II X-1 in 2009/2010 (MJD $\sim55200$) revealed variations in the X-ray luminosity by an order of magnitude on timescales of days to months (see Fig.~\ref{fig:ULX_LCs}). The 2009/2010 XRT monitoring observations also suggests that Ho II X-1 occupies either a high or low luminosity state. However, recent near-contemporaneous X-ray/mid-IR observations taken from 2017 - 2018 show Ho II X-1 exhibiting an intermediary X-ray flux between the high and low states observed in 2009/2010. Currently, the X-ray/mid-IR observations of Ho II X-1 do not reveal any clear evidence for correlated mid-IR and X-ray variations on short ($\sim$ day) or long ($\sim$years) timescales.

\subsubsection{NGC 7793 P13}

NGC 7793 P13, hereafter P13, is located in the spiral galaxy NGC 7793 (d = $3.4$ Mpc; Pietrzynski et al. 2010), exhibits X-ray luminosities up to L$_X\sim10^{40}$ erg s$^{-1}$, and is another one of the four known ULX pulsars (Furst et al. 2016; Israel et al. 2017b). P13 hosts a B9 supergiant mass donor and is one of only 2 ULXs that have a spectroscopically identified optical counterpart (Motch et al. 2014). Motch et al. (2014) presented UV and optical photometry of P13 that revealed a 64 day period, which agreed with the period they derived from radial velocity observations of a He II emission line. Furst et al. (2018) confirm a long orbital period of $63.9^{+0.5}_{-0.6}$ days of P13 from a timing analysis of multiple \textit{XMM-Newton} and \textit{NuSTAR} observations. 

Near- and mid-IR imaging observations of P13 with Magellan/FOURSTAR and \spitzer/IRAC reveal the first detections of an IR counterpart (Fig.~\ref{fig:ULX_dets} \&~\ref{fig:ULX_NIR}). $JHK_s$ imaging of P13 consistently show a an IR counterpart slightly elongated along the east-west direction. Based on $K_s$ imaging taken on 2011 September 8 (\ref{fig:ULX_NIR},\textit{right}), the minor axis of the IR counterpart is consistent with the average FWHM of point sources in the field of view ($0.9''$) and the FWHM of the major axis is $\sim20\%$ longer ($1.1''$). Assuming a distance of 3.4 Mpc to P13, the near-IR counterpart appears extended on size scales of $\lesssim 20$ pc. However, a plausible explanation for the elongation of the P13 near-IR counterpart is background source contamination from a collection of faint red stars. Notably, the optical $V$-band counterpart of P13 (Motch et al. 2014) does not appear elongated.

Due to the faintness of its IR emission and the crowding from background sources, multi-epoch \spitzer/IRAC observations provide only a few robust ($>3\sigma$) detections of the mid-IR counterpart (Fig.~\ref{fig:ULX_LCs}). The median absolute magnitudes of P13 are [3.6] = $-8.46\pm0.29$ and [4.5] = $-8.42\pm0.31$. No evidence of elongation is detected from the mid-IR counterpart, which is likely due to the limited angular resolution of Warm \spitzer/IRAC where the FWHM of the Warm Point Response Function (PRF) is $2.0''$. The mid-IR color of P13 is consistent with that of OB supergiants in the LMC, while the mid-IR brightness is comparable with the most mid-IR luminous B supergiants (Fig.~\ref{fig:ULX_CMD}). The IR + $V$ SED of P13 is consistent with the SED template of late B (8 and 9) supergiants within the $1\sigma$ uncertainties (Fig.~\ref{fig:ULX_SEDs}), which corroborates the B9 classification from Motch et al. (2014). However, the slightly rising slope of the near-IR SED and the enhanced mid-IR emission suggests the IR counterpart of P13 exhibits an IR excess over the expected late B supergiant emission. Unfortunately, it is difficult to distinguish whether or not the IR excess is due to an additional IR emitting component of P13 or a collection of faint red stars that may be contributing to its elongated near-IR appearance.



\subsubsection{NGC 253 ULX1}

NGC 253 ULX1, also known as RX J004722.4-252051, NGC 253 X9, and NGC 253 XMM2, is a ULX in the nearly edge-on spiral galaxy NGC 253 (d = $3.47\pm0.16$; Dalcanton et al. 2009) that exhibits an X-ray luminosity of $\sim3\times10^{39}$ ergs s$^{-1}$ (Barnard 2010). Barnard (2010) show from $\sim110$ ks \textit{XMM-Newton} observations that the X-ray emission from NGC 253 ULX1 in the 0.3 - 10 keV band is variable on the level of $\sim30\%$. Multi-epoch observations with \textit{Swift/XRT} show that NGC 253 ULX1 exhibits variability on $\sim$yr-timescales with observed maximum flux variations reaching a factor of $\sim4$ (Fig.~\ref{fig:ULX_LCs}).

H14 detected a near-IR counterpart of NGC 253 ULX1 with an absolute magnitude of $K_s = -10.5 \pm 0.5$, which they claim is consistent with an RSG. Near-IR imaging of NGC 253 ULX1 with Magellan/FourStar presented in this work measure a $K_s$ magnitude closely consistent with H14 (Tab.~\ref{tab:ULXNIRTab}). Follow-up near-IR spectroscopy of NGC 253 ULX1 by Heida et al. (2015) reveal CO and neutral metal absorption features, which confirm the RSG nature of the near-IR counterpart. 

\spitzer/IRAC observations of NGC 253 ULX1 detect a mid-IR counterpart with median absolute magnitudes and $3\sigma$ limits of [3.6] = $-10.03 \pm 0.26$ and [4.5] $>-9.79$. Although the mid-IR counterpart is present in the 4.5 $\mu$m observations based on visual inspection, it does not pass the $3\sigma$ detection threshold since the extended mid-IR emission from the galaxy and point sources in the vicinity of NGC 253 ULX1 complicate the background subtraction and raise the RMS flux estimate of the background. Due to the high photometric uncertainties due to the crowded background, it is unclear if the mid-IR counterpart of NGC 253 ULX1 is variable.

The 3.6 $\mu$m absolute magnitude and 4.5 $\mu$m limit of NGC 253 ULX1 are consistent with the RSGs on the LMC CMD (Fig.~\ref{fig:ULX_CMD}). Williams \& Bonanos (2016) arrive at a similar conclusion of the RSG-like color and magnitude of NGC 253 ULX1 in their independent analysis of \spitzer~photometry on point sources in nearby galaxies. The close agreement of the IR SED of NGC 253 ULX1 to the RSG template corroborates the RSG interpretation by H14 and Heida et al. (2015).


\subsubsection{Holmberg IX X-1}

Holmberg IX X-1, hereafter referred to as Ho IX X-1, is another well-studied ULX located in the dwarf galaxy Holmberg IX ($d = 3.61$ Mpc; Dalcanton et al. 2009). Ho IX X-1 exhibits a high X-ray luminosity of $L_X\sim10^{40}$ erg s$^{-1}$ (e.g. Kong et al. 2010), where the 0.3-10.0 keV flux has been observed to vary by a factor of $\sim3$ while the 15 - 40 keV flux varied by only $\sim20$ $\%$ (Walton et al. 2017). Ho IX X-1 is also surrounded by a $\sim300$ pc-sized nebula that is believed to be shock ionized (Pakull \& Mirioni 2002, Abolmasov \& Moiseev 2008). Previous near-IR $H$-band imaging by H14 did not detect a counterpart to Ho IX X-1 down to a $1\sigma$ limiting magnitude of $H > 20.25$, which corresponds to a limiting absolute magnitude of $>-8.0$. 

Archival \spitzer/IRAC observations of Ho IX X-1 published by Dudik et al.~(2016) revealed robust ($>10\sigma$) detections of $\sim10$ $\mu$Jy at both 3.6 and 4.5 $\mu$m in only three epochs taken between 2007 Nov 13 and Nov 15 (MJD 54417.9 - 54419.33). However, the mid-IR counterpart of Ho IX X-1 in 2007 Nov is not detected in the subsequent observation taken $\sim5$ months later in 2008 Apr 9 (MJD 54565.6) to a limiting $3\sigma$ flux of $\lesssim 3$ $\mu$Jy (see Fig.~\ref{fig:ULX_LCs}), which demonstrates the transient behavior of Ho IX X-1 in the mid-IR. Four follow-up \spitzer/IRAC observations taken between 2017 Jun 10 - Jun 16 (MJD 57914 - 57920) do not detect the mid-IR counterpart of Ho IX X-1, and near-contemporaneous \textit{Swift}/XRT observations consistently show the ULX to be in a low luminosity state (Fig.~\ref{fig:ULX_LCs}). There was, however, no X-ray coverage during the detection of mid-IR emission in 2007 Nov.

The limiting near and mid-IR absolute magnitudes of Ho IX X-1 in quiescence show that the mid-IR counterpart candidate is unlikely a RSG nor sgB[e] star (see Fig.~\ref{fig:ULX_SEDs}). 

\subsubsection{NGC 3031 ULX1}

NGC 3031 ULX1, also known as M81 X-6 and NGC 3031 X-11, is a ULX located in the nearby galaxy M81 (d = 3.61 Mpc; Durrell et al. 2010). NGC 3031 ULX1 has an average X-ray luminosity of $\sim2\times10^{39}$ ergs s$^{-1}$ and exhibits a factor of $40\%$ variability (Roberts \& Warwick 2000; Liu et al. 2002). NGC 3031 ULX1 is located near a large $\sim300$ pc shell-like nebula that is believed to be a supernova remnant (Pakull \& Mirioni 2003; Ramsey et al. 2006). Liu et al. (2002) identified an optical counterpart of NGC 3031 ULX1 with \textit{HST}/WFPC2 \textit{BVI} imaging observations where $V = 23.89 \pm 0.03$. They claim that the optical counterpart is a locally dust-obscured O8 V star with an extinction corrected absolute \textit{V}-band magnitude and $B-V$ color of $M_V = -4.9$ and $B-V$ = 0.32. H14 did not detect a near-IR counterpart down to a limiting $1\sigma$ magnitude of $K_s>18.5$, and near-IR follow-up with P200/WIRC from this work did not detect a near-IR counterpart down to limiting $3\sigma$ magnitudes of $K_s>18.85$ and $H>20.12$, or an absolute limiting magnitude of $K_s>-8.94$ and $H>-7.67$ (Tab.~\ref{tab:ULXNIRTab}). Based on the brightness of the optical counterpart reported by Liu et al. (2002), the near-IR observations were unlikely sensitive enough to detect a near-IR counterpart of an O8V photosphere.

\spitzer/IRAC observations of NGC 3031 ULX1 show that emission from its mid-IR counterpart is variable with 4.5 $\mu$m fluxes ranging from $<7$ $\mu$Jy to $18.7\pm1.8$ $\mu$Jy (Tab.~\ref{tab:ULXDetTab} and Fig.~\ref{fig:ULX_LCs}). The median 3.6 and 4.5 $\mu$m absolute magnitudes of NGC 3031 ULX1 for the epochs where it is detected are $-9.32 \pm0.4$ and $-9.73\pm0.31$, respectively. It is unclear if the optical counterpart is also variable, but the brightness and red color of the mid-IR counterpart suggests an excess of mid-IR emission. The SED of NGC 3031 ULX1 shows absolute mid-IR magnitudes consistent with sgB[e]s and RSGs, and its mid-IR color $[3.6]-[4.5] = 0.41\pm0.47$ places it in the color gap between red and blue ULXs. However, the non-supergiant O8 V star claim for the optical counterpart by Liu et al. (2002) cannot be ruled out since the mid-IR emission from NGC 3031 ULX1 was only detected in its brightest state while most of the observations were below the \spitzer~detection threshold. 



\subsubsection{M101 XMM1}


M101 XMM1, also known as J140314+541807 and NGC 5457 ULX2, is a ULX in the face-on spiral galaxy M101 (d = 6.43 Mpc; Shappee \& Stanek 2011) and exhibits an X-ray luminosity of $2.9\times10^{39}$ ergs s$^{-1}$ (Winter et al. 2006). H14 detected a near-IR counterpart and measured an absolute magnitude of $H = -10.69\pm0.1$ and claim it is consistent with an RSG. 

\spitzer/IRAC observations of M101 XMM1 measure median absolute magnitudes of [3.6] = $-11.16\pm0.07$ and [4.5] = $-11.27\pm0.09$ with small-amplitude variability on the order of $\sim20\%$ or $2\sigma$. This mid-IR variability is consistently measured at both 3.6 and 4.5 $\mu$m. The mid-IR properties of M101 XMM1 are therefore similar to that of M101 XMM3. The \spitzer~mid-IR photometry again supports the H14 hypothesis that the IR counterpart of M101 XMM1 is an RSG donor star.

\subsubsection{M101 XMM3}


M101 XMM3, also known as J1402+5440, NGC 5457 X23, and NGC 5457 ULX3, is another ULX in M101 and radiates with an X-ray luminosity of $2.4\times10^{39}$ ergs s$^{-1}$ (Swartz et al. 2011). H14 detected the near-IR counterpart of M101 XMM3 and measured an absolute magnitude of $H=-9.7\pm 0.2$, which they claim is consistent with a RSG. 

\spitzer/IRAC observations of M101 XMM3 provide median absolute magnitudes of [3.6] = $-11.21\pm0.07$ and [4.5] = $-11.33\pm0.09$ with small-amplitude variability on the order of $\sim10\%$\footnote{The \spitzer/IRAC observations on MJD 58323.95 shows a substantial increase in the 3.6 $\mu$m flux. A closer inspection of the image reveals a cosmic ray or hot pixel coincident with the source position. We therefore disregard this measurement.}. However, it is unclear if the variability is consistent at both 3.6 and 4.5 $\mu$m since M101 XMM3 was only observed with one channel each epoch due to its $\sim10'$ displacement from the center of the field of view. The mid-IR color and [3.6] absolute magnitude of M101 XMM3 are consistent with the brightest RSGs in the mid-IR (Fig.~\ref{fig:ULX_CMD}). The IR SED of M101 XMM3 also appears consistent with RSGs with the \spitzer~ photometry slightly above the upper $1\sigma$ end of the RSG template. The mid-IR photometry therefore support the hypothesis from H14 that the IR counterpart of M101 XMM3 is an RSG donor star.

\subsubsection{NGC 925 ULX1}

NGC 925 ULX1, also known as CXO J022727+333443, is  a ULX in the barred spiral galaxy NGC 925 ($d = 7.24\pm1.34$ Mpc; Tully et al. 2009) and exhibits one of the highest X-ray luminosities from a ULX $\sim2 - 4\times 10^{40}$ ergs s$^{-1}$ (Swartz et al. 2011; Pintore et al. 2018). Similar to the majority of ULXs with broadband X-ray coverage (e.g. Walton et al.~2018), \textit{Chandra} and \textit{XMM/NuSTAR} spectra of NGC 925 ULX1 by Pintore et al.~(2018) suggest the X-ray emission originates from super-Eddington accretion onto a stellar compact object as opposed to sub-Eddington accretion onto an IMBH. Pintore et al. (2018) also presented the \textit{Swift/XRT} observations of NGC 925 ULX1 that show fluctuations in the X-ray emission by factors of $\sim3$, which is demonstrated in Fig.~\ref{fig:ULX_LCs}. 

H14 detected a near-IR counterpart of NGC 925 ULX1 with absolute magnitudes of $H = -9.2\pm0.4$ and $K_s = -9.8 \pm 0.4$ and claim it is consistent with that of an RSG. Follow-up near-IR spectroscopy of NGC 925 ULX1 by Heida et al. (2016), however, revealed no detection of continuum emission and that the $H$-band spectrum is dominated by a single strong [Fe II] ($\lambda = 1.644$ $\mu$m) emission line with additional weaker H Brackett series, Helium, and H$_2$ emission lines. Heida et al. (2016) interpret the presence of these low ionization lines as evidence for dense circumstellar material around the ULX. Pintore et al. (2018) also suggest the presence of surrounding nebulosity based on observations of diffuse H$\alpha$ around the source.

\spitzer/IRAC observations of NGC 925 ULX1 reveal a mid-IR counterpart with median absolute magnitudes of [3.6] = $-11.34\pm0.42$ and [4.5] = $-12.13\pm0.41$. The mid-IR light curve shows small-amplitude $\sim10\%$ variability at both 3.6 and 4.5 $\mu$m (Fig.~\ref{fig:ULX_LCs}), which is $\sim2\sigma$ from the median brightness. The mid-IR flux and red colors of NGC 925 ULX1 are consistent with the sgB[e]s shown in the LMC CMD (Fig.~\ref{fig:ULX_CMD}). The mid-IR emission in the IR SED of NGC 925 ULX1 is also consistent with the sgB[e] template. The near-IR photometry from H14 slightly exceeds the upper $1\sigma$ end of the sgB[e] template.
It is unclear what the nature of the mid-IR candidate counterpart of NGC 925 ULX1 is, but the red mid-IR excess suggests the presence of dust around the ULX.

\subsubsection{NGC 925 ULX2}


NGC 925 ULX2, also known as CXO J022721+333500, is another ULX in NGC 925 that was observed in excess of $10^{39}$ ergs s$^{-1}$ (Swartz et al. 2011). \textit{Chandra} and \textit{XMM/NuSTAR} observations presented by Pintore et al. (2018) show that NGC 925 ULX2 has a constant X-ray light curve, which is consistent with the constant \textit{Swift/XRT} light curve shown in Fig.~\ref{fig:ULX_LCs}.

H14 detected a slightly extended near-IR counterpart to NGC 925 ULX2 with absolute magnitudes of $H = 10.6 \pm 0.4$ and $K_s = -11.3 \pm0.4$, which they suggest may be associated with multiple RSGs. Follow-up near-IR spectroscopy of NGC 925 ULX2 by Hedia et al. (2016) revealed neutral metal and CO absorption features that are consistent with an RSG.

\spitzer/IRAC observations detect a mid-IR counterpart with median absolute magnitudes of [3.6] = $-11.74\pm0.47$ and [4.5] = $-11.87 \pm 0.47$ with no significant variability throughout archival, SPIRITS, and short-cadence observations (Fig.~\ref{fig:ULX_LCs}). The mid-IR brightness and blue colors of NGC 925 ULX2 are consistent with the RSGs shown in the LMC CMD (Fig.~\ref{fig:ULX_CMD}). The IR SED of NGC 925 ULX2 exhibits consistently high emission in excess of $1\sigma$ brighter than the median RSG emission (Fig.~\ref{fig:ULX_SEDs}). The overall IR spectral shape of NGC 925 ULX2 resembles than of an RSG, which suggests that the IR counterpart may indeed be composed of multiple RSGs. One of these RSGs is therefore likely to be the ULX donor star.


\subsubsection{NGC 4631 X4}

NGC 4631 X4, also known as CXO J124157.4+323202, is a transient ULX in the barred spiral galaxy NGC 4631 (d = 7.35 $\pm0.74$ Mpc; Tully et al. 2013). Soria \& Ghosh (2009) revealed that NGC 4631 X4 transitioned in state between the X-ray observations taken by \textit{Chandra} on 2000 Apr 16 and \textit{XMM-Newton} on 2002 Jun 28. The \textit{Chandra} observations detected NGC 4631 X4 as a faint, soft source with an emitted luminosity of $L_X\sim3\times10^{37}$ ergs s$^{-1}$, whereas the \textit{XMM-Newton} observations detected a brighter, harder source with an unabsorbed luminosity of $L_X\sim2\times10^{39}$ ergs s$^{-1}$ (Soria \& Ghosh 2009). NGC 4631 X4 has therefore existed in a sub-Eddington non-ULX state, and was classified as a high-mass X-ray binary (HMXB) by Mineo, Gilfanov, \& Sunyaev (2012) based on its \textit{Chandra}-measured luminosity.


L17 do not detect a near-IR counterpart to NGC 4631 X4 down to a $1\sigma$ limiting absolute magnitude of $H>-10.03$, and P200/WIRC observations from this work do not detect it down to a $3\sigma$ limiting magnitude of $J > -8.72$, $H>-9.28$, and $K_s>$-10.22. Unlike the other mid-IR counterparts that exhibit variability, the \spitzer/IRAC light curve of NGC 4631 X4 shows tentative evidence of periodic variability where the flux ranges from $\sim 30 - 60$ $\mu$Jy at 3.6 and 4.5 $\mu$m (Fig.~\ref{fig:ULX_LCs}). Two mid-IR flux peaks at MJD$\sim57100$ and $\sim57900$ seen at both 3.6 and 4.5 $\mu$m are separated by $\sim800$ days. Assuming a mid-IR variable period of 800 days, NGC 4631 X4 should have exhibited a mid-IR peak at MJD$\sim53100$. This is consistent with the timing of the first \spitzer/IRAC observations of NGC 4631 X4 taken around MJD$=53150$, which exhibits fluxes at the same level as the mid-IR peaks.

The median mid-IR absolute magnitudes of NGC 4631 X4 are [3.6] = $-12.21\pm0.28$ and [4.5] = $-12.88\pm0.24$, which indicate a red color consistent with sgB[e] stars and a brightness consistent with the most mid-IR luminous sgB[e] stars in the LMC (Fig.~\ref{fig:ULX_CMD}). The IR SED of NGC 4631 X4 (Fig.~\ref{fig:ULX_SEDs}) shows that the \spitzer~absolute magnitudes are around the upper $1\sigma$ end of the mid-IR flux from sgB[e]s and the near-IR limiting absolute magnitudes disfavor an RSG interpretation. 





\subsubsection{NGC 4136 ULX2}

NGC 4136 ULX2, also known as J120922+295559, is a ULX located in the spiral galaxy NGC 4136 (d = 9.55 Mpc; Tully 1988) and exhibits an X-ray luminosity of $2.6\times 10^{39}$ ergs s$^{-1}$ (Roberts et al. 2004). H14 detected a near-IR counterpart with an absolute magnitude of $H = -10.75 \pm 0.4$ and claim it is consistent with an RSG. Follow-up near-IR spectroscopy of NGC 4136 ULX2 by Heida et al. (2016) revealed neutral metal and CO absorption lines consistent with an RSG photosphere; however, the most prominent features were [Fe II] ($\lambda = 1.644$ $\mu$m) and Hydrogen Brackett series emission lines, which suggest the presence of a nebula. 

Mid-IR observations with \spitzer/IRAC of NGC 4136 ULX2 measure median absolute magnitudes of [3.6] = $-12.5 \pm 0.41$ and [4.5] = $-12.49 \pm 0.41$. NGC 4136 was not observed in SPIRITS, therefore there are only archival observations taken 2004 May 27 (MJD 53152.7) and the 4 observations taken our short-cadence mid-IR study taken between 2017 September 15 - 22 (MJD 58011 - 58018). Neither observations reveal significant mid-IR variability from NGC 4136 ULX2 (Fig.~\ref{fig:ULX_LCs}). The mid-IR color and magnitude of NGC 4136 ULX2 is consistent with the brightest RSGs in the LMC (Fig.~\ref{fig:ULX_CMD}). A comparison of NGC 4136 ULX2 with the IR SED templates shows that the mid-IR flux is $\sim3\sigma$ brighter than the median RSG flux, and the near-IR $H$-band photometry from H14 are consistent with the upper $1\sigma$ end of the RSG flux when accounting for uncertainties. The excess mid-IR emission may arise from dust in the vicinity of the ULX or the counterpart candidate may be composed of multiple RSGs like the interpretation of NGC 925 ULX2.

\begin{figure}[t]
	\centerline{\includegraphics[width=1\linewidth]{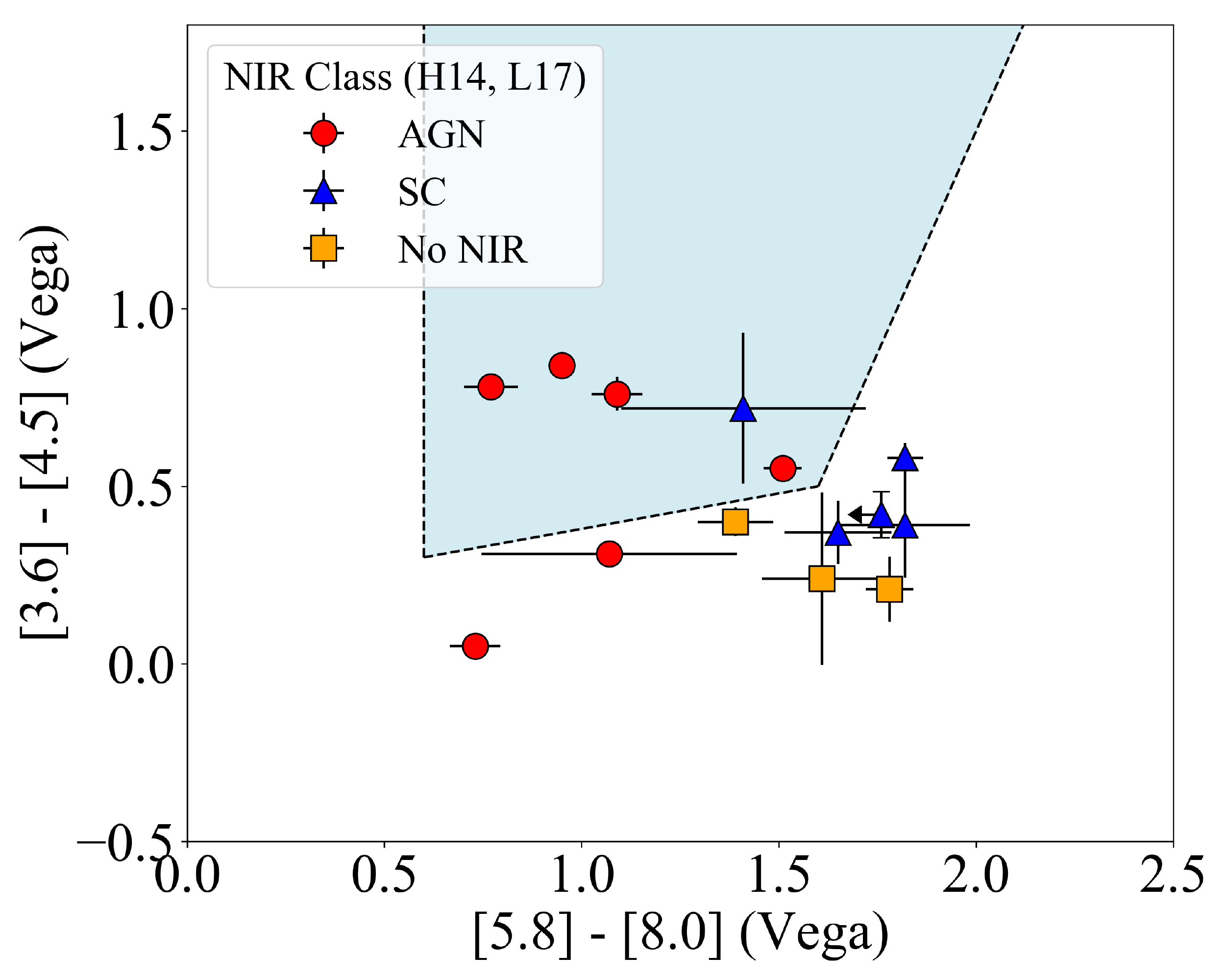}}
	\caption{\spitzer/IRAC color-color diagram of 14/\ODULX of the star cluster, AGN, or background galaxy candidates plotted over a shaded region corresponding to empirically derived color-color region of active galaxies (Stern et al.~2005). The red circles and blue triangles indicate AGN and star cluster classifications based on near-IR photometry by H14 and L17. Orange squares correspond to sources that did not have any near-IR counterpart detected by H14 or L17.}
	\label{fig:ULX_CCD}
\end{figure}

\subsection{Background AGN or Star Clusters Candidates} 
\label{sec:ULX_AGN}

\ODULX~of the \NULX~ULXs in this search for \spitzer/IRAC mid-IR ULX counterparts were detected with absolute magnitudes brighter than expected for supergiants: [3.6] $< -12.5$ and [4.5] $<-13.0$ (Fig.~\ref{fig:ULX_ODets}). These mid-IR counterpart candidates are likely associated with star clusters in the host galaxy or background AGN. A background AGN counterpart therefore implies that the associated X-ray emission does not originate from a ULX. 

13 of the \ODULX~sources were detected in the near-IR by H14 and L17 and given preliminary classifications based on their near-IR absolute magnitudes, proximity to the apparent host galaxy, or mid-IR colors from the Wide-Field Infrared Survey Explorer (WISE, Wright et al. 2010). We compare these NIR classifications with the results from a mid-IR color-color analysis using the empirically derived \spitzer/IRAC colors of active galaxies vs Galactic stars and normal galaxies by Stern et al. (2005). The \spitzer~color-color diagram is shown in Figure~\ref{fig:ULX_CCD} with 14 of the \ODULX~mid-IR counterparts\footnote{Only Ch2 and Ch4 \spitzer/IRAC observations of the source 1WGA J0940.0+7106 in Holmberg I were found in the Spitzer Heritage Archive, and NGC 4258 X9 did not have any archival Ch3 and Ch4 Cold \spitzer/IRAC observations}.

\subsubsection{1WGA J0940.0+7106 in Holmberg I}

1WGA J0940.0+7106 is the only one of the bright mid-IR detections that does not have \spitzer/IRAC imaging coverage at 3.6 and 5.8 $\mu$m and so is not included in Fig.~\ref{fig:ULX_CCD}. At a distance of Holmberg I, the mid-IR brightness from Cold \spitzer~observations is [4.5] = $-13.00\pm 0.03$, which only slightly exceeds the range of mid-IR supergiant emission. H14 reported a counterpart with a near-IR absolute magnitude consistent with a single RSG; however, they note that the counterpart is located 5' away from the center of Holmberg I, which has a diameter of 3.6' (Gil de Paz et al. 2007). H14 also show that the WISE-measured mid-IR colors are consistent with an AGN. Despite nearly consistent IR magnitudes with supergiants, the large separation from the host galaxy and the WISE colors indicate that the counterpart is most likely a background AGN.

\subsubsection{[LB2005] NGC 5457 X26 in M101}
The mid-IR colors of NGC 5457 X26 are [3.6] - [4.5] = $ 0.72 \pm 0.21 $ and [5.8] - [8.0] = $ 1.41 \pm 0.31 $, which is consistent with the AGN region in the color-color diagram (Fig.~\ref{fig:ULX_CCD}). However, three other bright mid-IR knots and diffuse extended emission are located in the vicinity of the counterpart (Fig.~\ref{fig:ULX_ODets}). At a distance of M101, the absolute magnitudes of this source from Cold \spitzer~observation are [3.6] = $ -13.49 \pm 0.17 $ and [4.5] = $ -14.21 \pm 0.14 $. L17 detected a bright near-IR counterpart and suggested it could be an unresolved young star cluster. The position of the IR counterpart is consistent with a bright X-ray knot located in the giant HII region NGC 5471 located within M101 (Skillman 1985). Based on a \textit{Chandra} X-ray analysis of NGC 5471, Sun et al. (2012) classify this knot, referred to as NGC 5471B, as a ``hypernova remnant" candidate. The discrepant mid-IR colors of the IR counterpart from the other candidate star clusters may reflect its different emission properties as a hypernova remnant.

\subsubsection{2XMM J140248.0+541350 in M101}

The mid-IR colors of 2XMM J140248.0+541350 are [3.6] - [4.5] = $ 0.55 \pm 0.02 $ and [5.8] - [8.0] = $ 1.51 \pm 0.05 $, which are consistent with that of an AGN. At a distance of M101, the absolute magnitude of this source from Cold \spitzer~observations is [3.6] = $ -13.26 \pm 0.05 $ and [4.5] = $ -13.81 \pm 0.05 $. The mid-IR emission of this source is close to one magnitude above the emission from supergiants. H14 detected an extended near-IR counterpart with a similar excess in emission compared to RSGs and suggest it is a background AGN. The mid-IR colors and the bright IR emission support the counterpart's identification as a background AGN.

\subsubsection{[LB2005] NGC 5457 X32 in M101}
The mid-IR colors of NGC 5457 X32 are [3.6] - [4.5] = $ 0.05 \pm 0.01 $ and [5.8] - [8.0] = $ 0.73 \pm 0.06 $, which is significantly outside of the typical color-color region of AGN. At a distance of NGC 5457, the absolute magnitudes of this source from Cold \spitzer~observations are [3.6] = $ -14.44 \pm 0.05 $ and [4.5] = $ -14.49 \pm 0.05 $. The mid-IR emission is therefore brighter than a supergiant by $\sim2$ magnitudes and also appears spatially extended (Fig~\ref{fig:ULX_ODets}). L17 detected a counterpart with a near-IR absolute magnitude consistent with that of a single RSG; however, they find that the near-IR counterpart is spatially extended but not located near a star-forming region and is therefore likely the host galaxy of a background AGN. The IR counterpart is spatially coincident with an optically identified galaxy in the Sloan Digital Sky Survey (SDSS) J140134.57+542031.2 (Adelman-McCarthy et al. 2009). Although it falls outside of the AGN region in the mid-IR color-color space, the IR counterpart is most likely a background AGN.

\subsubsection{RX J121845.6+472420 in NGC 4258}

The mid-IR colors of J121845.6+472420 are [3.6] - [4.5] = $ 0.4 \pm 0.04 $ and [5.8] - [8.0] = $ 1.39 \pm 0.10 $, which is just outside of the AGN region in the color-color diagram (Fig.~\ref{fig:ULX_CCD}). It is located 6.4' from the nucleus of its apparent host galaxy NGC 4258 and appears coincident with one of its spiral arms. At a distance of NGC 4258, the absolute magnitudes of this source from Cold \spitzer~observations are [3.6] = $ -12.37 \pm 0.11 $ and [4.5] = $ -12.77 \pm 0.11 $. Although the 3.6 and 4.5 $\mu$m absolute magnitudes are consistent with the emission from a supergiant, the mid-IR counterpart is slightly spatially extended and exhibits an excess at longer IR wavelengths: [5.8] = $ -13.73 \pm 0.13 $ and [8.0] = $ -15.12 \pm 0.12 $. Similar to J115733.7+552711 in NGC 3990, a mid-IR counterpart is detectable out to $\sim20$ $\mu$m with \spitzer/MIPS and WISE W4. Based on a WISE color-color analysis, W1 - W2 = 0.3 and W2 - W3 = 4.3, the mid-IR counterpart is located far off from the color-color region occupied by AGN (Mateos et al. 2012). H14 do not detect a near-IR counterpart down to a limiting magnitude of $H>21.0$, or an absolute magnitude of $H> -8.3$.

A spectral and timing analysis of XMM-Newton observations of NGC 4258 by Akyuz et al. (2013) suggested that the X-ray counterpart, referred to as NGC 4258 XMM-10, could be an X-ray binary in NGC 4258 with an unabsorbed X-ray luminosity of $9.2\times10^{38}$ erg s$^{-1}$. Akyuz et al. (2013) also find that the X-ray source's short-term X-ray light curve ($<1$ day) does not exhibit variability, but has increased by a factor of at least 5 over a decade.

Based on its coincidence with a spiral of NGC 4258 and the suggested X-ray binary classification from Akyuz et al.~(2013), we interpret the mid-IR counterpart of J121845.6+472420 as a stellar cluster. The lack of a near-IR counterpart and the increasing mid-IR excess towards longer wavelengths implies the cluster may be enshrouded locally by gas and dust.





\subsubsection{RX J121844.0+471730 in NGC 4258}

The mid-IR colors of RX J121844.0+471730 are [3.6] - [4.5] = $ 0.37 \pm 0.09 $ and [5.8] - [8.0] = $ 1.65 \pm 0.14$, which falls outside of the AGN region in the color-color diagram (Fig.~\ref{fig:ULX_CCD}). At a distance of NGC 4258, the absolute magnitudes of this source from Cold \spitzer~observations are [3.6] = $ -13.58 \pm 0.13 $ and [4.5] = $ -13.95 \pm 0.12$. At the center of a cluster of stars, H14 detected a bright near-IR counterpart exceeding the emission of a single RSG. The mid-IR counterpart is also located near a collection of mid-IR sources (Fig.~\ref{fig:ULX_ODets}). The counterpart's IR brightness, mid-IR color, and proximity to additional sources indicates it is part of a stellar cluster.

\subsubsection{RX J121857.7+471558 in NGC 4258}

The candidate mid-IR counterpart of RX J121857.7+471558, also known as NGC 4258 X3 (Liu \& Mirabel 2005), appears to be located in one of the spiral arms of NGC 4258 and is at the edge of the 1.2''-radius error circle centered on its X-ray position. The mid-IR counterpart is slightly spatially extended (FWHM$=3.1$'') and its mid-IR colors are [3.6] - [4.5] = $ 0.21 \pm 0.09 $ and [5.8] - [8.0] = $ 1.78 \pm 0.06 $, which are outside of the AGN region of the color-color diagram but consistent with the 4 of the 5 star cluster candidates (Fig.~\ref{fig:ULX_CCD}). At a distance of NGC 4258, the absolute magnitudes from Cold \spitzer~observations are [3.6] = $ -13.52 \pm 0.13 $ and [4.5] = $ -13.73 \pm 0.13$. H14 do not report a detection of a near-IR counterpart down to a limiting magnitude of $H>19.0$, or an absolute magnitude of $H>-10.3$, which suggests the region is highly obscured. Given its apparent location in the spiral arm of its host galaxy and the consistent colors with the other star cluster candidates, the mid-IR counterpart NGC 4258 X3 is likely a star cluster. 

\subsubsection{[LB2005] NGC 4258 X9 in NGC 4258}

NGC 4258 X9 did not have any archival Cold \spitzer/IRAC images, but a candidate mid-IR counterpart appears as a bright point-like source in Warm \spitzer/IRAC observations and exhibits bright mid-IR absolute magnitudes of [3.6] = $ -15.83 \pm 0.11 $ and [4.5] = $ -16.50 \pm 0.11 $. L17 reported a bright near-IR counterpart and characterized it as an AGN based on its WISE colors. The WISE mid-IR counterpart is also classified as an AGN in Secrest et al.~(2015); therefore, the mid-IR counterpart is likely an AGN.


\subsubsection{[SST2011] J124211.13+323235.9 in NGC 4631}

The mid-IR colors of J124211.13+323235.9 are [3.6] - [4.5] = $ 0.39 \pm 0.15 $ and [5.8] - [8.0] = $ 1.82 \pm 0.16$, which falls outside of the AGN region in the color-color diagram (Fig.~\ref{fig:ULX_CCD}). At a distance of NGC 4631, the absolute magnitudes of this source from Cold \spitzer~observations are [3.6] = $ -16.53 \pm 0.24 $ and [4.5] = $ -16.92 \pm 0.24 $. The mid-IR counterpart is significantly brighter than a supergiant and is located in a crowded region consistent with the disk plane of NGC 4631. L17 detected a bright near-IR counterpart that also appears extended and suggest that it is a star cluster. Based on the mid-IR colors, the IR brightness, its extended appearance, and location within the apparent host galaxy, the IR counterpart is most likely a star cluster.

\subsubsection{[SST2011] J123029.55+413927.6 in NGC 4490}

The mid-IR colors of J123029.55+413927.6 are [3.6] - [4.5] = $ 0.58 \pm 0.04 $ and [5.8] - [8.0] = $ 1.82 \pm 0.05 $, which falls outside of the AGN region in the color-color diagram (Fig.~\ref{fig:ULX_CCD}). This ULX appears to be located in the pair of interacting galaxies NGC 4490/NGC 4485 (e.g. Roberts et al. 2002). At a distance of NGC 4490, the absolute magnitude of this source from Cold \spitzer~observations is [3.6] = $ -16.07 \pm 0.18 $ and [4.5] = $ -16.65 \pm 0.17$. The mid-IR emission from the counterpart is significantly brighter than a supergiant and also appears spatially extended. L17 detected an extended and bright near-IR counterpart and suggested it is a star cluster. The counterpart is also located in the western arm of NGC 4490 which is linked by a ``bridge'' of material to a tidal tail south of NGC 4485 where enhanced star formation may be occurring (Roberts et al. 2002). The IR counterpart is therefore most likely a stellar cluster.

\subsubsection{XMMU J132953.3+471040 in M51}

The mid-IR colors of XMMU J132953.3+471040 are [3.6] - [4.5] = $ 0.42 \pm 0.07 $ and [5.8] - [8.0] = $<1.76$. At a distance of M51, the absolute magnitudes of this source from Cold \spitzer~observations are [3.6] = $ -15.33 \pm 0.21 $ and [4.5] = $ -15.75 \pm 0.21 $. The spatially extended mid-IR counterpart is coincident with the spiral arm of M51 and is surrounded by diffuse galaxy emission and mid-IR point sources. H14 detected a bright and extended near-IR counterpart that appears to be associated with a star cluster shown in  imaging from the \textit{Hubble Space Telescope} (Terashima et al. 2006). The counterpart is likely associated with a star cluster within M51.

\subsubsection{XMMU J024323.5+372038 in NGC 1058}

The mid-IR colors of XMMU J024323.5+372038 are [3.6] - [4.5] = $0.78\pm0.03$ and [5.8] - [8.0] = $0.77\pm0.07$, which are consistent with the color-color region occupied by AGN (Fig.~\ref{fig:ULX_CCD}). At a distance of NGC 1058, the absolute magnitudes of this source from Cold \spitzer~observations are [3.6] = $-12.97\pm0.4$ and [4.5] = $-13.75\pm0.4$. This is within a magnitude above the mid-IR emission expected for supergiants; however, the mid-IR emission appears partially extended at 3.6 and 4.5 $\mu$m (Fig.~\ref{fig:ULX_ODets}). H14 reported a point-source counterpart with near-IR magnitudes consistent with that of an RSG but note that the source is located in the outskirts of NGC 1058 in a region with no signs of recent star formation. H14 therefore suggest it is a foreground star or background AGN. The mid-IR colors and extended mid-IR appearance of XMMU J024323.5+372038 indicate that the counterpart is likely a background AGN. 

\subsubsection{[IWL2003] 68 in NGC 1637}

The mid-IR colors of [IWL2003] 68 are [3.6] - [4.5] = $0.76\pm0.04$ and [5.8] - [8.0] = $1.09\pm0.06$, which place it robustly in the color-color region occupied by AGN (Fig.~\ref{fig:ULX_CCD}). At a distance of NGC 1637, the absolute magnitudes of this source from Cold \spitzer~observations are [3.6] = $-14.71\pm0.4$ and [4.5] = $-15.47\pm0.4$. This mid-IR emission is more than 2 magnitudes brighter than that expected from a supergiant. H14 reported a spatially resolved near-IR $K_s$-band counterpart of [IWL2003] 68 with an absolute magnitude far exceeding that of a single RSG and suggest the counterpart is either a star cluster or AGN. The mid-IR colors and bright mid-IR flux measured by \spitzer~indicate that this counterpart, and thus the ULX, is likely a background AGN.

\subsubsection{3XMM J115733.7+552711 in NGC 3990}

The candidate mid-IR counterpart of J115733.7+552711 is located on the outskirts of NGC 3990, $\sim0.4$ arcminutes from its core. The mid-IR colors of the counterpart are [3.6] - [4.5] = $ 0.24 \pm 0.24 $ and [5.8] - [8.0] = $ 1.61 \pm 0.15 $, which falls outside of the AGN region in the color-color diagram (Fig.~\ref{fig:ULX_CCD}). At a distance of NGC 3990, the absolute magnitudes of this source from Cold \spitzer~observations are [3.6] = $ -13.6 \pm 0.35 $ and [4.5] = $ -13.84 \pm 0.35 $.  Despite its brightness in the mid-IR, L17 do not detect a near-IR counterpart down to a limiting magnitude of $H>19.11$, or an absolute magnitude of $H>-10.71$. There is no optical counterpart in SDSS nor PanSTARRS catalogs. 

Interestingly, the mid-IR counterpart is detectable out to longer IR wavelengths with the Multi-Band Imaging Photometer (MIPS; Rieke et al. 2004) on \spitzer~at 24 $\mu$m and WISE W4 (22 $\mu$m). Based on a WISE color-color analysis, W1 - W2 = 0.1 and W2 - W3 = 2.3, the mid-IR counterpart is located far off from the color-color region occupied by AGN (Mateos et al. 2012). This is consistent with the \spitzer~color-color analysis. However, given its IR brightness, the counterpart may be an obscured background AGN.



\subsubsection{CXO J080157.8+504339 in NGC 2500}

The mid-IR colors of CXO J080157.8+504339 are [3.6] - [4.5] = $0.84\pm0.01$ and [5.8] - [8.0] = $0.95\pm0.02$, which are consistent with the color-color region occupied by AGN (Fig.~\ref{fig:ULX_CCD}). At a distance of NGC 2500, the absolute magnitudes of this source from Cold \spitzer~observations are [3.6] = $-15.93\pm0.17$ and [4.5] = $-16.77\pm0.17$. The mid-IR emission far exceeds the expected emission from supergiants. H14 also reported a near-IR counterpart much brighter than expected for an RSG. Optical spectroscopy by Gutierrez (2013) indicate that the counterpart is a background AGN, which corroborates the AGN-like mid-IR colors.

\subsubsection{[LB2005] NGC 4594 X5 in NGC 4594}

The mid-IR colors of NGC 4594 X5 are [3.6] - [4.5] = $ 0.31 \pm 0.03$ and [5.8] - [8.0] = $ 1.07 \pm 0.32 $, which falls slightly outside of the AGN color-color space (Fig.~\ref{fig:ULX_CCD}). At a distance of NGC 4594, the absolute magnitudes of this source from Cold \spitzer~observations are [3.6] = $ -13.78 \pm 0.26 $ and [4.5] = $ -14.09 \pm 0.26 $. L17 detected a near-IR counterpart with an absolute magnitude of $H = -11.61\pm 0.57$, consistent with a single RSG within uncertainties. Based on its WISE color, L17 suggest it is consistent with an AGN (e.g. Mateos et al. 2012). 

The WISE All-Sky Data Release catalog (Cutri et al. 2012) provides mid-IR measurements at W1 (3.35 $\mu$m) and W2 (4.6 $\mu$m) of 16.81 $\pm0.15$ and 15.41 $\pm0.12$, respectively. The WISE color W1 - W2 = 1.4 is significantly redder than the [3.6] - [4.5] \spitzer/IRAC color of the mid-IR counterpart. Since the \spitzer/IRAC observations were taken between mid-2004 and early-2005 (MJD 53166 - 53392), where $[4.5] = 16.17 \pm 0.03$, and the WISE observations were taken between early- and mid-2010 (MJD 55204 - 55379), it is plausible that the mid-IR counterpart exhibited significant fluctuations if it is indeed an AGN. The mid-IR counterpart is therefore likely a background AGN. 



\subsection{Mid-IR Non-Detections with Near-IR Counterparts} 

Of the \NDULX~ULXs that do not have an apparent mid-IR counterpart, \NIRMULX~have near-IR counterparts detected by H14 and L17 with imaging observations from LIRIS/WHT or SWIRC/MMT. 9 of these near-IR counterparts were suggested to be single RSG donor stars: J073655.7+653542, J110545.62+000016.2, J121847.6+472054, J123030.82+413911.5, J123558+27577, J132947+47096, J112018.32+125900.8, J120922.6+295551, and NGC 5408 X-1. The 2 remaining near-IR counterparts with non detections in the mid-IR were suggested to be multiple RSGs (J123038.4+413831, L17) and a star cluster (J123043.1+413819, L17). The \spitzer/IRAC $3\sigma$ limiting magnitudes of these sources range from $\sim -10.5$ to $\sim-13.5$ (Tab.~\ref{tab:ULXNonDetTab}) due to the crowding and high background emission. Mid-IR limits on the IR counterparts are consistently brighter than the near-IR absolute magnitudes and therefore do not conflict with the near-IR counterpart candidate classifications by H14 and L17.

\section{Discussion}

\subsection{Evidence for Variable Jet Activity from Holmberg IX X-1}
\label{Sec:HoIX}
The mid-IR counterpart of Ho IX X-1 exhibited the most dramatic variable behavior of the mid-IR counterparts (Fig.~\ref{fig:ULX_LCs}). Based on the \spitzer~observations taken when Ho IX X-1 was in outburst, Dudik et al.~(2016) interpreted the origin of the mid-IR emission as dust emission in a circumbinary disk or a variable jet. Dudik et al.~(2016), however, only analyzed the \spitzer~observations of Ho IX X-1 when it was detected on 2007 Nov 15. The transient mid-IR behavior of Ho IX X-1 supports the interpretation of a variable jet as the origin of the mid-IR counterpart detected in 2007 Nov. 

With near-simultaneous Cold \spitzer/IRAC observations of Ho IX X-1 at 3.6, 4.5, 5.8, and 8.0 $\mu$m (Ch1 - 4), we can estimate the spectral index, $\alpha$, of the mid-IR emission, where $F_\nu\propto \nu^\alpha$. During peak mid-IR flux on 2007 Nov 13 (MJD 54418), the 3.6 and 4.5 $\mu$m measurements indicate a relatively flat index of $\alpha=-0.19\pm0.1$. Based on the 3.6 and 4.5 $\mu$m fluxes, this spectral index is consistent with the $3\sigma$ non-detection limits at 8.0 $\mu$m. Although there is a 5.8 $\mu$m detection of the Ho IX X-1 mid-IR counterpart, there is substantial contamination from nearby nebulosity that falls in the aperture so it is difficult to resolve the 5.8 $\mu$m flux contribution from the ULX counterpart itself. A flat or slightly inverted index with a ``jet break" in the IR has been observed from inferred jet activity of black hole X-ray binaries such as GX339-4 (Gandhi et al.~2011) and V404 Cyg (Russell et al.~2013). This evidence supports the interpretation of the Ho IX X-1 mid-IR counterpart detected in 2007 Nov as a variable jet.


\subsection{Distinguishing Emission from Dust, Donor star, and Accretion Disk}
\label{Sec:DDD}
ULX optical counterparts are difficult to distinguish between the photospheres of hot donor-stars, accretion disk outflows. For example, optical spectroscopy of SS 433 and CXOU J140332.3+542103\footnote{This ULX is in our mid-IR sample but is not detected by \spitzer/IRAC.} (a.k.a. M101 ULX-1) both resemble a WR star, but in SS 433 it is the accretion disk outflow that mimics a WR appearance (van den Heuvel 1981; Fuchs et al.~2006) whereas M101 ULX-1 hosts a WR mass-donor (Liu et al.~2013). Optical counterparts may also arise from X-ray irradiated accretion disks (e.g. Tao et al.~2011), and should exhibit a power-law in the optical SED and variability correlated with the X-ray emission. 

Emission from sources resembling O or B-star photospheres peaks at optical/UV wavelengths but their contribution to the IR decreases by over an order of magnitude. At IR wavelengths, dominant emitting components from ULXs may be red (supergiant) donor-stars, surrounding dust, free-free emission from accretion disk winds, or non-thermal jet emission: e.g. NGC 925 ULX2 (Heida et al.~2016), Ho II X-1 (Lau et al.~2017), SS 433 (Fuchs et al.~2006), and Ho IX X-1 (this work), respectively. However, in the case of Ho IX X-1, non-thermal jet emission in the mid-IR exhibits high amplitude variability on short ($\lesssim$ month) timescales. Since all the other stellar-like ULXs do not exhibit such transient behavior (Fig.~\ref{fig:ULX_LCs}), we do not associate their mid-IR emission with jet activity. 

The spectral profiles of free-free emission, RSG photospheres, and thermal dust emission can be differentiated by their IR SEDs (e.g. Bonanos et al.~2009). We note that with the sensitivity depth of our sample we are unlikely to detect mid-IR counterparts dominated by free-free emission. For example, the ``red" appearance of WR stars in the \spitzer~CMD (open circles, Fig.~\ref{fig:ULX_CMD}) is due to free-free emission, which only approaches M$_\mathrm{[3.6]}<-8$ (Bonanos et al.~2009). SS 433, whose IR emission arises from free-free emission, exhibits an absolute 3.6 $\mu$m magnitude of M$_\mathrm{[3.6]}\sim-7.6$ adopting $F_{3.6}\sim1$ Jy and $d = 5.5$ kpc (Fuchs et al.~2006). Therefore, besides Ho IX X-1, we claim that the stellar-like mid-IR ULX counterparts are dominated by the donor star or surrounding dust. The ``blue" ULXs host RSG donor stars while the ``red" ULXs have circumstellar/binary dust.

We note that blue ULXs with IR counterparts resembling RSG donor stars can also exhibit optical counterparts due to the outflows from the accretion disk and/or X-ray irradiated outer disk. If NGC 300 ULX1 is an RSG, those sources may be origin of the \textit{g}-band excess (Fig.~\ref{fig:ULX_SEDs}).

\subsection{Circumstellar/binary Dust around ULXs}
\label{sec:dust}

\begin{figure*}[t!]
	\centerline{\includegraphics[width=1.0\linewidth]{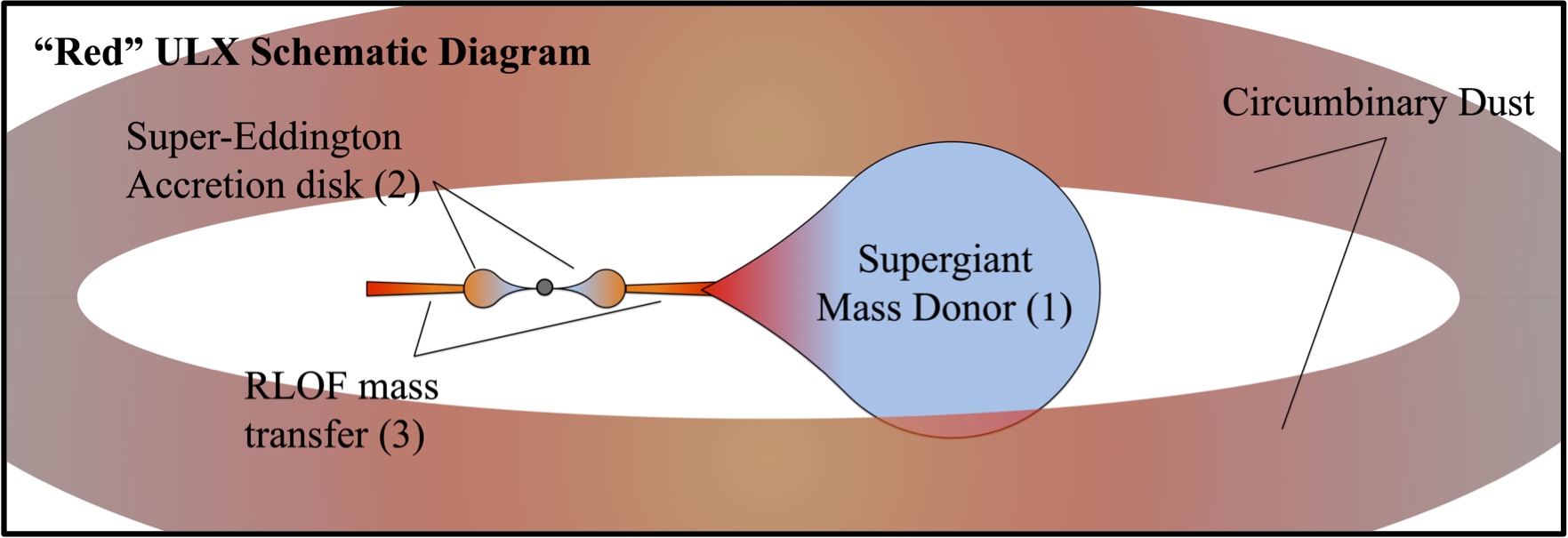}}
	\caption{``Red" ULX schematic illustration (not to scale) showing circumbinary dust surronding a supergiant mass donor undergoing RLOF and transferring mass to a compact object accreting at Super-Eddington rates. The numbers (1 - 3) correspond to the different mass-loss/outflow scenarios considered in Sec.~\ref{sec:ULXBe} that may give rise to the ``B[e] phenomenon": (1) outflows from the supergiant donor star, (2) winds from the geometrically and optically thick Super-Eddington accretion disk, and (3) mass-loss from the outer Lagrangian points from RLOF mass transfer.}
	\label{fig:RedULX}
\end{figure*}

In previous work on the \spitzer~mid-IR counterpart of Holmberg II X-1, we argued that the mid-IR emission is dominated by hot dust ($\sim600-800$ K; Lau et al. 2017). In this subsection, we discuss the derivation of dust properties using Ho II X-1 assuming radiative heating from the donor star and apply the calculations to the other red ULXs in Table~\ref{tab:ULXDustTab}. A schematic illustration of a red ULX is shown in Fig.~\ref{fig:RedULX}. We address the dust heating contribution from an isotropic X-ray radiation field (Pakull \& Mirioni 2003; Kaaret et al. 2004; Berghea et al. 2010) and whether or not it will destroy dust by heating it beyond its sublimation temperature of $\sim1700$ K in the following subsections~\ref{sec:dust2} \&~\ref{sec:dust3}, respectively.


The 3.6 and 4.5 $\mu$m flux measurements from the mid-IR counterparts can be used to derive properties of dust that dominates the IR emission. Thermal dust emission as a function of wavelength can be expressed as $F_\lambda \propto Q_\mathrm{em}(\lambda,a )\,B_\lambda(T_d)$, where $F_\lambda$ is the flux at wavelength $\lambda$, $Q_\mathrm{em}(\lambda,a)$ is the grain emissivity model for dust grains of radius $a$, and $B_\lambda(T_d)$ is the Planck function for a dust temperature, $T_\mathrm{d}$. Dust temperatures can then be derived from the 3.6 and 4.5 $\mu$m flux ratio

\beq
\frac{F_{3.6}}{F_{4.5}}=\frac{Q_\mathrm{em}(3.6\,\mu\mathrm{m},a )\,B_{3.6}(T_d)}{Q_\mathrm{em}(4.5\,\mu\mathrm{m},a )\,B_{4.5}(T_d)}.
\label{eq:Td}
\eeq


\noindent Assuming a single temperature component of optically thin silicate grains with $a=0.1$ (Laor \& Draine 1993) and using the median 3.6 and 4.5 $\mu$m fluxes from Holmberg II X-1, we find that the median dust temperature is $T_\mathrm{d}=720^{+140}_{-100}$ K. 

The total IR dust luminosity, $L_\mathrm{IR}$, can be estimated by integrating of the dust emission model normalized to the 4.5 $\mu$m flux and adopting a distance, $d$, to the source:

\beq
L_\mathrm{IR}= 4 \pi d^2 \int F_{4.5}\frac{ Q_\mathrm{em}(\lambda,a )  B_{\lambda} (T_d)}{Q_\mathrm{em}(4.5\,\mu\mathrm{m},a )  B_{4.5} (T_d)}\,\mathrm{d}\lambda.
\label{eq:Lum2}
\eeq



\noindent Using the median mid-IR flux measurements from Holmberg II X-1, we derive an IR luminosity of $L_\mathrm{IR}=2.6^{+0.8}_{-0.4}\times10^4$ $\mathrm{L}_\odot$.

The mass of the emitting dust, $M_\mathrm{d}$, can be estimated from the mid-IR flux and dust model as shown in Eq.~\ref{eq:Mass}:

\beq
M_d=\frac{(4/3)\,a\,\rho_b\,F_\lambda\,d^2}{Q_\mathrm{em}(\lambda,a )\,B_\lambda(T_d)},
\label{eq:Mass}
\eeq

\noindent
where $\rho_b$ is the bulk density of the dust grains and $d$ is the distance to the source. A bulk density of $\rho_b=3$ $\mathrm{gm}$ $\mathrm{cm}^{-3}$ (e.g. Draine \& Li 2007) is assumed for the emitting silicate grains. The dust mass estimated from the median mid-IR flux and dust temperature of the Ho II X-1 counterpart is $M_\mathrm{d}=1.1^{+1.2}_{-0.6}\times10^{-6}$ $\mathrm{M}_\odot$.

Under the assumption that dust heating is balanced by thermal radiative cooling, the distance between the heating source and the emitting dust can be determined. This equilibrium temperature radius, $R_\mathrm{eq}$, is calculated from the equilibrium/steady-state dust temperature, $T_\mathrm{eq}$, and radiation fields from the donor star and the ULX:

\beq
R_\mathrm{eq}=\left(\frac{1}{16\pi}\frac{\int Q_{em}(\lambda,a)\,(L_*(\lambda)+L_{X}(\lambda))\,\mathrm{d}\lambda}{\int Q_{em}(\lambda,a)\,\pi B_\lambda(T_{eq})\,\mathrm{d}\lambda}\right)^{1/2},
\label{eq:temp1}
\eeq

\noindent where $L_*(\lambda)$ and $L_X(\lambda)$ are the stellar and ULX radiative heating components, respectively. Eq.~\ref{eq:temp1} also assumes that $R_\mathrm{eq}$ is much greater than the separation between the donor star and ULX.

Following the formalism of Draine (2011), we can define the spectrum-averaged absorption cross-section of each heating source, $\left<Q\right>_{*/X}$,\footnote{$\left<Q\right>_{*}$ and $\left<Q\right>_{X}$ are the spectrum-averaged absorption cross-sections for a stellar and X-ray heating source, respectively} and the Planck-averaged emission efficiency of the dust, $\left<Q\right>_{IR}$, as

\beq
\begin{split}
\left<Q\right>_{*/X} = \frac{\int Q_{em}(\lambda,a)\,L_{*/X}(\lambda)\,\mathrm{d}\lambda}{\int L_{*/X}(\lambda)\,\mathrm{d}\lambda} \\
\left<Q\right>_{IR} = \frac{\int Q_{em}(\lambda,a)\,\pi B_\lambda(T_{eq})\,\mathrm{d}\lambda}{\int\pi B_\lambda(T_{eq})\,\mathrm{d}\lambda}.
\end{split}
\label{eq:Qs}
\eeq

\noindent This simplifies Eq.~\ref{eq:temp1} to

\beq
R_\mathrm{eq}=\left(\frac{\left<Q\right>_* L_* + \left<Q\right>_X L_X}{\left<Q\right>_{IR}\,16\pi\sigma T^4_{eq}}\right)^{1/2}.
\label{eq:temp2}
\eeq

\noindent For now, we neglect the heating contribution from the ULX, $\left<Q\right>_X L_X$, and assume that the radiative dust heating is dominated by the stellar component. For $T_{eq} = T_d =720$ K and a stellar 20,000 K heating source with $L_* = L_{IR} = 2\times10^4$ L$_\odot$, we find $\left<Q\right>_*/ \left<Q\right>_{IR} \sim 30$ and

\begin{equation}
R_\mathrm{eq}^\mathrm{IR} \sim 130\, \mathrm{AU}.
\label{eq:ReqIR}
\end{equation}

\noindent This equilibrium temperature radius calculation is almost an order of magnitude larger than that derived in Lau et al.~(2017) for Ho II X-1. The reason for this discrepancy is a smaller value adopted for $\left<Q\right>_*/ \left<Q\right>_{IR} \sim 0.3$, which was based on a dust sublimation radius analysis by Smith et al. (2016). The low value of $\left<Q\right>_*/ \left<Q\right>_{IR} \sim 0.3$ quoted in Smith et al.~(2016) was later confirmed to be a typographical error thus motivating our full treatment of determining $\left<Q\right>_*/ \left<Q\right>_{IR}$ in this work.

The assumption of $L_\mathrm{IR}=L_*$ may under-predict the heating source luminosity since the thermal emission re-radiated by dust is dependent on the dust coverage fraction and geometry around the heating source. There may also be an additional heating compontent from the ULX X-ray emission (see Sec.~\ref{sec:dust2}). Therefore, we treat $R_\mathrm{eq}^\mathrm{IR}$ as a lower limit on the separation between the dust and the heating source.

Based on \textit{HST} imaging observations, Kaaret et al.~(2004) modeled the photoionizing radiation field of the optical counterpart of Ho II X-1 as an O5V star with $T_* = 42000$ K and $L_* = 8.3\times10^5$ L$_\odot$ and as a B2Ib star with $T_* = 18500$ K and $L_* = 5.2\times 10^4$ L$_\odot$. Assuming the O5V and B2Ib heating source, we find 

\begin{equation}
    \left<Q\right>_{O5V}/ \left<Q\right>_{IR} \sim \left<Q\right>_{B2Ib}/ \left<Q\right>_{IR} \sim 30,
\label{eq:Q*}
\end{equation}

from which it follows that

\begin{equation}
R_\mathrm{eq}^\mathrm{O5V} \sim 750\, \mathrm{AU} \,\,\mathrm{ and }\,\, R_\mathrm{eq}^\mathrm{B2Ib} \sim 190\, \mathrm{AU}. 
\end{equation}

\noindent Dust around Ho II X-1 should therefore be located in the range of $\sim130 - 750$ AU from a stellar heating source. The unresolved $K_s$ band emission centered on Ho II X-1 shown in Fig.~\ref{fig:HoII_FC} (\textit{right}), which likely arises from dust (See Fig.~2 in Lau et al.~2017b),  suggests that dust must be located within $<0.16''$, or $<2.6$ pc, of the ULX as opposed to the surrounding $\sim20$ pc nebula.



\subsection{X-ray Dust-Heating Contribution}
\label{sec:dust2}

X-ray ionized nebulae observed around ULXs indicate an isotropic radiation field emitted form the ULX (e.g. Pakull \& Mirioni 2003). In the case of Ho II X-1, the nebula requires an X-ray ionizing flux of $>4\times10^{39}$ erg s$^{-1}$  (Kaaret et al. 2004), which implies that a significant fraction of the X-ray luminosity ($L_X\sim10^{40}$ erg s$^{-1}$) measured from Ho II X-1 is is radiated isotropically and may therefore contribute to heating circumstellar/binary dust. 

\begin{figure}[t]
	\centerline{\includegraphics[width=1.0\linewidth]{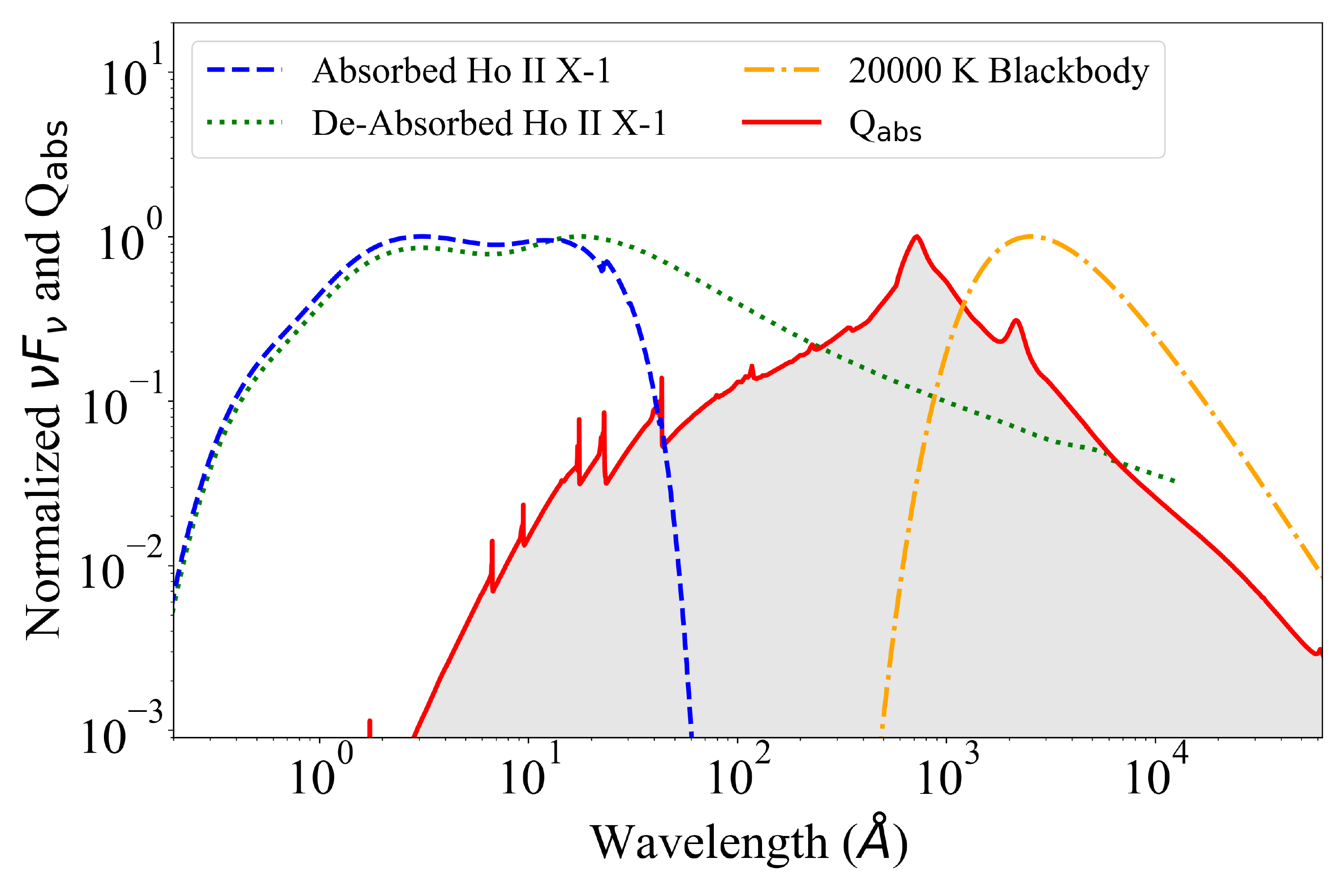}}
	\caption{Normalized dust absorption efficiency, Q$_\mathrm{abs}$ (red solid line), from Weingartner \& Draine (2001), Li \& Draine (2001) and Draine (2003a, b, \& c) overlaid with the normalized absorbed (blue dashed line) and de-absorbed (green dotted line) Ho II X-1 from Walton et al. (2018) and a 20,000 K blackbody.}
	\label{fig:HoIIRF}
\end{figure}

Although the total observed X-ray luminosity from a ULX is greater than the luminosity of a supergiant, $L_X > L_*$, the dust heating efficiency of X-ray photons is more than an order of magnitude lower than optical/UV photons (e.g. Weingartner \& Draine 2001). We use the X-ray radiation field of Ho II X-1 to assess the radiative dust-heating contribution relative to the stellar component. For this ULX dust heating calculation, we adopted the Weingartner \& Draine (2001), Li \& Draine (2001), and Draine (2003a, b, \& c) absorption models\footnote{\url{ftp://ftp.astro.princeton.edu/draine/dust/mix/kext_albedo_WD_MW_3.1_60_D03.all}} for interstellar dust composed of a mixture of carbonaceous and amorphous silicate grains and is valid for photon energies as high as 12.4 keV. Figure~\ref{fig:HoIIRF} shows the normalized dust absorption efficiency, Q$_\mathrm{abs}$, overlaid with a 20,000 K blackbody, which represents the photosphere of an early B-type star, and the X-ray radiation field of Ho II X-1. For the X-ray radiation field, we adopted the absorbed X-ray emission model of Ho II X-1 by Walton et al.~(2018), which has a 0.3 - 12.4 keV X-ray luminosity of $7.7\times10^{39}$ erg s$^{-1}$. It follows from Eq.~\ref{eq:Qs} that $\left<Q\right>_{X}/ \left<Q\right>_{IR} \sim 1$, which implies that the relative dust-heating efficiency from a stellar B2Ib-like source (Eq.~\ref{eq:Q*}) and a Ho II X-1-like ULX is

\beq
\frac{\left<Q\right>_{*}}{\left<Q\right>_{X}}\sim 30. 
\eeq

\noindent Therefore, for an early-B or late-O stellar heating source with a supergiant luminosity of $L_*\sim10^5$ L$_\odot$ and a ULX with an X-ray luminosity of $L_X\sim10^{40}$ erg s$^{-1}$, the radiative dust-heating input is comparable:

\beq
L_*\left<Q\right>_{*}\sim L_X \left<Q\right>_{X}.
\eeq

Since the addition of the X-ray heating component increases the value of $R_\mathrm{eq}^{IR}$ (e.g. Eq.~\ref{eq:ReqIR}) by a factor of $\sim\sqrt{2}$, it is valid to assume that the $R_\mathrm{eq}^{IR}$ derived from only a stellar heating component with luminosity $L_\mathrm{IR}$ provides a lower limit on the location of dust around the ULX system. 

Another possible dust-heating component could be the reprocessed optical/UV emission from an X-ray irradiated outer accretion disks of the ULX (e.g. Tao et al.~2011). However, $\sim0.1\%$ of the X-ray emission is found to be reprocessed to optical/UV wavelengths for ULXs (Tao et al.~2011); therefore, the IR luminosities we derive for the red ULXs (Tab.~\ref{tab:ULXDustTab}) exceed the possible radiative input from outer accretion disks by over an order of magnitude. Thus, we do not expect irradiated accretion disks to significant contribute to heating surrounding dust for the red ULXs in our sample. 

\subsection{Dust Evaporation by X-Rays?}
\label{sec:dust3}

One potential issue regarding the existence of dust in the vicinity of a ULX is the exposure to this hard and luminous X-ray radiation field. Given the X-ray luminosity of a ULX and assuming that $\left<Q\right>_{X}/ \left<Q\right>_{IR} \sim 1$, as shown above for Ho II X-1, we can estimate the distance out to which dust would reach its sublimation temperature of $T_\mathrm{sub} = 1700$ K ( Eq.~\ref{eq:temp2}). For Ho II X-1, which has a 0.3 - 12.4 keV X-ray luminosity of $7.7\times10^{39}$ erg s$^{-1}$, dust will evaporate around the ULX out to distances of

\begin{equation}
R_\mathrm{sub}\sim 40\, \mathrm{AU}.
\end{equation}

\noindent This indicates that dust can indeed survive exposure to the Ho II X-1 radiation field at the $\sim130 - 800$ AU distances previously derived. 


The absorbed X-ray emission model for Ho II X-1 may underestimate the luminosity that surrounding dust is exposed to. 
To test dust survival in the extreme case that it is exposed to the full radiation field from the ULX, a \textit{de-absorbed} Ho II X-1 X-ray emission model by Walton et al.~(2018; Fig.~\ref{fig:HoIIRF}) where the 0.001 - 12.4 keV (1 - 1.2$\times10^4$ $\AA$) X-ray luminosity is $1.3\times10^{40}$ erg s$^{-1}$ is adopted. This model assumes a super-Eddington disc profile extrapolated to low energies, which may actually over-predict the low energy/long wavelength emission. The sublimation radius with this de-absorbed model is $R_\mathrm{sub}\sim 120$ AU.
As expected, this sublimation radius is larger than that derived from the absorbed X-ray emission model since softer X-ray photons are present in the de-absorbed model. However, the de-absorbed emission model sublimation radius is still smaller than the estimated dust distances, which reinforces the claim that dust can survive direct exposure to the Ho II X-1 radiation field.
In Tab.~\ref{tab:ULXDustTab}, $R_\mathrm{sub}$ is estimated for each of the red ULXs given their observed X-ray luminosities and assuming $\left<Q\right>_{X}/ \left<Q\right>_{IR} = 1$.






\subsection{Red ULXs and the B[e] Phenomenon}
\label{sec:ULXBe}

The 4 ULXs that exhibit ``red'' \spitzer~mid-IR counterparts and appear consistent with sgB[e] stars are Holmberg II X-1, NGC 925 ULX1, NGC 4631 X4, and pre-2010 outburst NGC 300 ULX1 (Fig.~\ref{fig:ULX_CMD}~\&~\ref{fig:ULX_SEDs}). The mid-IR excess emission and analysis in Sec.~\ref{sec:dust} suggest the counterparts have a hot ($\sim500 - 1000$ K) circumstellar dust component, which is one of the signature characteristics of the ``B[e] phenomenon'' (Lamers et al. 1998). Optical or near-IR spectroscopy show forbidden emission lines of [Fe II] and strong Hydrogen emission lines from Holmberg II X-1, NGC 925 ULX1 and NGC 300 ULX1 (Heida et al. 2016, Villar et al. 2016), which are also defining characteristics of the B[e] phenomenon\footnote{One of the defining properties of the ``B[e] phenomenon" is strong \textit{Balmer} emission lines, which are most likely present given the strength of the near-IR H Brackett lines shown by Heida et al. (2016).}. We therefore claim that these 4 mid-IR counterparts are ULXs exhibiting the B[e] phenomenon and suggest three possible mass-loss scenarios (Fig.~\ref{fig:RedULX}) that could be producing it:


1. The mass donor in the ULX is a sgB[e] star and the B[e] phenomenon arises from its outflow, which is similar to the interpretation of sgB[e] mass donors in the wind-fed Galactic HMXBs CI Cam (Clark et al. 1999, Bartlett et al.~2018) and IGR J16318-4848 (Filliatre \& Chaty 2004). 

2. The B[e] phenomenon arises from an outflow from an optically and geometrically thick accretion disk around the compact object in the ULX. Disk winds have been inferred from optical spectroscopy of ULXs (e.g. Fabrika et al. 2015) and 
have also been attributed to enhanced equatorial outflows from the Galactic super-Eddington accreting X-ray binary SS 433 (Blundell et al. 2001; Fabrika 2004; Fuchs et al. 2006).

3. Mass-loss through the outer Lagrangian point L2 behind the compact object from non-conservative mass transfer during Roche-Lobe Overflow (RLOF) produces an enhanced equatorial outflow and presents the B[e] phenomenon (e.g. Zickgraf 2003; Miroshnichenko 2007; Clark et al. 2013). Observations of the B[e] phenomenon seen from the galactic supergiant B[e] colliding-wind binary Wd1-9 are interpreted to arise from rapid mass-transfer via RLOF (Clark et al.~2013).


We first consider scenario 1, where the mass donor is a sgB[e] star. Lau et al. (2017) proposed this interpretation for Ho II X-1 based on the comparable mid-IR colors and IR luminosity of known sgB[e] stars. The stellar companions in several galactic HMXBs have also been interpreted as sgB[e] stars (e.g. Filliatre \& Chaty 2004), which suggests an sgB[e]-ULX is plausible. However, the difference between the HMXB and ULX scenarios is that the donor star in a ULX is likely undergoing RLOF, which should affect the mass-loss from a sgB[e] donor if it is driven by rapid rotation.

The enhanced mass-loss from sgB[e]s are believed to be driven by rapid stellar rotation at $>50\%$ of its break-up velocity. However, if the sgB[e] is undergoing RLOF to feed the compact object, the sgB[e] will lose angular momentum and become tidally locked to the orbit, where its rotation period will be the same as the orbital period. We can estimate the expected rotational velocity of a tidally locked sgB[e] given the Roche lobe radius expression derived by Eggleton (1983) where Roche lobe radius of the donor star, $R_L$, can be related to $a$ and the mass ratio of the donor and accretor, $q$:


\beq
\frac{R_L}{a} = \frac{0.49\,q^{2/3}}{0.6\,q^{2/3}+\mathrm{Log}(1+q^{1/3})}.
\label{eq:a}
\eeq
\noindent
Assuming a sgB[e] mass and radius of M$_*=30$ M$_\odot$ and R$_*=27$ R$_\odot$, consistent with the sgB[e] star in IGR J16318-4848 (Filliatre \& Chaty 2004; Rahoui et al. 2008), a compact object accretor mass of 20 M$_\odot$ black hole, we find that when $R_L=R_*$ the rotational velocity of the sgB[e] star, $V_{rot}$ would be 

\beq
v_{rot}\sim25\,\mathrm{km s}^{-1} \sim 5\%\, v_{crit},
\label{eq:b}
\eeq
\noindent
where we have applied Kepler's Third Law after deriving $a$ from Eq.~\ref{eq:a}, and $v_{crit}\approx \sqrt{GM_*/R_*}$. We note that $v_{rot}$ does not have a strong dependence on the mass of the accretor. For example, if M$_{BH}=5$ M$_\odot$, then $v_{rot}\sim30$ km s$^{-1}$. It is therefore unlikely that the B[e] phenomenon in ULXs originate from an sgB[e] mass donor star since in RLOF it should not be rotating at a significant enough fraction of their break-up velocity to produce an enhanced equatorial outflow.
However, we note that stellar ``wind RLOF" has recently been proposed by El Mellah, Sundqvist, \& Keppens (2019) to provide enough mass to feed ULX compact objects at super-Eddington rates and thus do not entirely rule out the possibility of scenario 1.

We now consider scenario 2, where the B[e] phenomenon arises from a dense outflow from the outer parts of the ULX accretion disk. Optical spectroscopy of ULXs show that their accretion disks drive winds and exhibit similar physical conditions as massive and evolved stars like late-type nitrogen-rich Wolf-Rayet (WNL) stars and LBVs (Fabrika et al.~2015), the latter of which is spectroscopically similar to sgB[e] stars (e.g. Humphreys et al. 2017). This accretion disk outflow scenario has been proposed for the galactic super-Eddington accreting binary SS 433 (Fuchs et al. 2006), where its IR spectra resemble a WNL star. Excess mid-IR emission indicate the presence of circumstellar dust around SS 433, the source of which Fuchs et al. (2006) argue is its accretion disk winds. Radio observations of SS 433 also show an extended ``ruff'' along the equatorial plane extending out to distances $>100$ AU (Blundell et al. 2001), which verifies the presence of an equatorial outflow. The ULX accretion disk winds from red, dusty ULXs may therefore be mimicking a sgB[e] star and producing the B[e] phenomenon.

Another mechanism for dust formation around stellar binaries requires colliding winds that produce regions of high densities that can cool and form dust (Usov 1991). This dust production mechanism manifests as elegant spiral or arc-like dust plumes for a class of late-type carbon-rich Wolf-Rayet stars (WCL) with OB star companions (e.g. Monnier, Tuthill \& Danchi 1999, 2002; Tuthill et al.~1999; Williams et al. 2009). Since the dense outflows from ULX accretion disks can mimic Wolf-Rayet (WR) photospheres and are in close orbits with likely supergiant mass donors (e.g. Motch et al. 2014), we propose that dust formation in ULXs may also occur via colliding winds.

In scenario 3, we suggest the B[e] phenomenon may arise from equatorial mass-loss through the outer Lagrangian points from mass transfer during RLOF from the donor star onto the compact object in the ULX. Such binary interaction processes have been proposed for producing the B[e] phenomenon (Zickgraf 2003; Miroshnichenko 2007; Clark et al. 2013) and is another interpretation for the enhanced equatorial mass-loss and circumstellar dust formation around SS 433 (e.g. Fabrika 1993; Paragi et al. 1999; Clark et al. 2007). Dust has been observed to form in this manner in the Galactic blue supergiant eclipsing binary RY Scuti, which is currently undergoing binary mass transfer via RLOF (Smith et al.~2011). With our observations it is difficult to distinguish between scenarios 2 and 3. We therefore conclude that the B[e] phenomenon in the red ULXs most likely arises from either scenario 2 or 3 (or a combination of the two scenarios), but does not arise from a sgB[e] mass donor undergoing RLOF as proposed in scenario 1.

\subsection{Red ULX Mass Donors and Mass-loss Rates}

In our proposed picture of red ULXs (Fig.~\ref{fig:RedULX}), the IR counterpart is dominated by emission from surrounding dust and therefore does not correspond to a direct detection of the mass donor star. Thus, in scenarios 2 and 3, the nature of the mass donor is not constrained to be a sgB[e] star. However, red ULX mass donors are likely O or B-type main sequence or supergiant stars given the high luminosities $\sim(3 - 17)\times10^4$ L$_\odot$ re-radiated by dust in the IR (Tab.~\ref{tab:ULXDustTab}), where it is assumed that the mass donor is the primary radiative heating source of surrounding dust. Red ULX mass donors are unlikely to be RSGs since RSGs would exhibit excess near-IR emission between $\sim1 - 2$ $\mu$m (Fig.~\ref{fig:ULX_SEDs}).

Interestingly, observations of the post-2010 outburst ``color gap" ULX NGC 300 ULX1 (Fig.~\ref{fig:ULX_CMD}) show such a near-IR excess that appears consistent with an RSG while the mid-IR photometry is more consistent with circumstellar dust emission from sgB[e] stars. This may suggest that the mass donor of the post-2010 outburst NGC 300 ULX1 is an RSG while the sgB[e]-like mid-IR emission arises from dust production via scenarios 2 and/or 3. 

The presence of circumstellar/binary dust provides a diagnostic for estimating mass-loss rates of the dust-forming outflow from red ULXs. This mass-loss rate can be approximated as the total outflow mass traced by emitting dust divided by the elapsed time for this material to travel from the central system to its apparent location:

\begin{equation}
    \dot{M}\sim\frac{M_d\,X_{g/d}}{R_\mathrm{eq}\,v_w^{-1}}.
\label{eq:ULX_Mdot}
\end{equation}
\noindent
In Eq.~\ref{eq:ULX_Mdot}, $M_d$ is the observed dust mass, $X_{g/d}$ is the gas-to-dust mass ratio, $R_{eq}$ is the equilibrium temperature radius of the surrounding dust, and $v_w$ is the velocity of the winds from which dust is formed. For the gas-to-dust mass ratio we adopt the canonical value of $X_{g/d}=100$, and we assume a wind velocity of $\sim400$ km s$^{-1}$ which is consistent with the outflows observed in the dusty sgB[e]-HMXBs IGR J16318-4848 and CI Cam (Coleiro et al.~2013; Filliatre \& Chaty 2004, Clark et al.~1999).  Given the consistent dust mass and equilibrium temperature radii exhibited by the 4 red ULXs of $M_d\approx(1-5)\times10^{-6}$ M$_\odot$ and $R_{eq}\approx 100 - 300$ AU, respectively, we adopt values of $M_d=3\times10^{-6}$ M$_\odot$ and $R_{eq} = 200$ AU. From Eq.~\ref{eq:ULX_Mdot}, we then estimate a mass-loss rate of

\begin{equation}
\begin{split}
    \dot{M}\sim 1\times10^{-4}\left(\frac{M_d}{3\times10^{-6}\,\mathrm{M}_\odot}\right)\left(\frac{X_{g/d}}{100}\right) \\ \left(\frac{200\,\mathrm{AU}}{R_\mathrm{eq}}\right) \left(\frac{v_w}{400\,\mathrm{km}\,\mathrm{s}^{-1}}\right)\,\mathrm{M}_\odot\,\mathrm{yr}^{-1}.
\end{split}
\label{eq:ULX_Mdot2}
\end{equation}
\noindent
We therefore find that the red ULX mass-loss rate of the dust-forming outflow is consistent with the mass-loss rate from the optically thick accretion disk outflow around SS433 ($\dot{M}\sim2-3\times10^{-4}$ M$_\odot$ yr$^{-1}$; Fuchs et al.~2006). The similarity to SS433 reinforces our ULX dust formation/mass-loss scenarios 2 and/or 3 (Sec.~\ref{sec:ULXBe}), both of which were interpreted as mass-loss mechanisms for SS433.

\section{Conclusions}

We performed a time-domain search for mid-IR counterparts of ULXs in nearby galaxies with \spitzer/IRAC.  Of the \NULX~ULXs in our sample, \NIRULX~exhibited mid-IR counterparts consistent with stellar supergiant-like fluxes, and \ODULX~had mid-IR counterparts with fluxes exceeding that of supergiants and were more consistent with background galaxies/AGN, or star clusters. \NDULX~out of the \NULX~ULXs did not have a detectable mid-IR counterpart.

Most of the stellar candidate mid-IR counterparts exhibit a dichotomy in their [3.6] - [4.5] colors, where four appear ``red'' ([3.6] - [4.5] $\sim0.7$) and five appear ``blue'' ([3.6] - [4.5] $\sim0$). Based on a \spitzer/IRAC CMD of supergiants in the LMC (Bonanos et al.~2009), we found that blue and red ULXs have absolute mid-IR magnitudes and colors consistent with RSG and sgB[e] stars, respectively (Fig.~\ref{fig:ULX_CMD}). A near- to mid-IR SED analysis of these ULX counterparts in comparison to supergiant ``template" SEDs support the CMD-based identifications.

The three other ULXs from the stellar candidate sample that were not included in these 9 red and blue ULXs were NGC 7793 P13, NGC 3031 ULX1, and Holmberg IX X-1. The mid-IR counterpart of P13 is fainter than the RSG-like ULXs but is consistent with the mid-IR color and magnitude of late B supergiants, consistent with its optical spectroscopic classification (Motch et al.~2014). The uncertainties of the mid-IR color of NGC 3031 ULX1 were too large to place in either red or blue ULX group. Ho IX X-1 exhibited a mid-IR flux consistent with a stellar counterpart but only in two epochs in 2007 Nov (Fig.~\ref{fig:ULX_LCs}), with only non-detections in subsequent and previous observations.

We investigated the mid-IR variability from Ho IX X-1, where the counterpart exhibited a relatively flat spectral index of $\alpha = -0.19\pm0.1$  and claim that the transient mid-IR detection of Ho IX X-1 in 2007 Nov was likely due to a variable jet, which was one of the two hypotheses proposed by Dudik et al.~(2016).


Of the \ODULX~mid-IR ULX counterparts with absolute magnitudes exceeding that of supergiants, 13 had near-IR counterparts with preliminary classification by H14 and/or L17 as star clusters or non-ULX background AGN. We performed a color-color analysis with archival Cold \spitzer/IRAC observations taken at 3.6, 4.5, 5.8, and 8.0 $\mu$m of the counterparts and tested the classifications using empirically derived AGN color-color regions defined by Stern et al.~(2005; Fig.~\ref{fig:ULX_CCD}). 
\NDULX~ ULXs do not have a detected \spitzer/IRAC mid-IR counterpart. The average $3\sigma$ limiting mid-IR absolute magnitudes on these observations were $-11.6\pm1.9$ and $-11.8\pm1.8$ for [3.6] and [4.5], respectively.

We discussed the advantage of distinguishing the emitting ULX counterparts in the IR and claim that all of the stellar-like mid-IR counterparts besides Ho IX X-1 are dominated by emission from either the donor star or surrounding dust.
We claim that the mid-IR emission from the blue ULXs are due to an RSG donor star, where as a red ULX counterpart arises from thermal emission from circumstellar/binary dust. 
An analysis of the X-ray radiation field from Holmberg II X-1 suggests that X-rays may contribute to heating dust but will not evaporate dust grains at the estimated equilibrium termperature radii (Tab.~\ref{tab:ULXDustTab}).

The circumstellar dust and observed strong Hydrogen and/or [Fe II] emission lines from these red ULXs are distinguishing characteristics of the ``B[e] phenomenon.'' It is unlikely that the B[e] phenomenon arises from a true sgB[e] mass donor since RLOF would slow the rotation of the sgB[e] companion to velocities below the threshold required to produce the dense equatorial outflow attributed with the phenomenon (e.g. Zickgraf 2003). However, the donor stars are likely O or B main sequence or supergiant stars given the luminosity re-radiated in the IR. We propose that the B[e] phenomenon is due to either an outflow from the outer parts of the ULX accretion disk around the compact object or mass-loss through the outer Lagrangian point during non-conservation mass-transfer in RLOF. 



By using the dust around red ULXs as a probe for the surrounding total mass, we estimated mass-loss rates of $\dot{M}\sim\,1\times10^{-4}\,\mathrm{M}_\odot\,\mathrm{yr}^{-1}$ in the dust-forming ULX system outflow. This is consistent with the optically thick accretion disk outflow around SS 433 ($\dot{M}\sim2-3\times10^{-4}$ M$_\odot$ yr$^{-1}$; Fuchs et al.~2006).

In this study, we demonstrated that the relatively unexplored utility of mid-IR observations to uniquely probe on the nature of ULXs. Characterizing the energetics and mass of surrounding dust in the red ULXs provides an opportunity to investigate the radiation field and mass-loss in in super-Eddington accreting compact objects. Although we are currently limited by the achievable spatial resolution for space-based mid-IR observations, upcoming facilities such as the 6.5-m \textit{James Webb Space Telescope} will open new windows to explore the mass donors and circumstellar environments of ULXs. 


\emph{Acknowledgments}. 
We thank the anonymous referee for their valuable comments and insight that improved the quality of this work.
We also acknowledge C.~Alvarez, N.~Blagorodnova, and the staff of the W. M. Keck Observatory for supporting observations and the assistance with data reduction. We also thank M.~Bachetti, B.~Binder, H.~Bond, Y.~G\"{o}tberg, B.~Grefenstette, I.~El Mellah, P.~Jonker, P.~Kosec, E.~Levesque, K.~Lopez, C.~Pinto, P.~Podsiadlowski, and R.~Soria for the enlightening discussions on ULXs, RSGs, sgB[e]s, and dust.
RML acknowledges the Japan Aerospace Exploration Agency’s International Top Young Fellowship. DJW acknowledges financial support from STFC in the form of an Ernest Rutherford fellowship. RDG was supported by NASA and the United States Air Force. JJ is supported by the National Science Foundation Graduate Research Fellowship under Grant No. DGE-1144469.
This work is based in part on observations made with the \textit{Spitzer} Space Telescope, which is operated by the Jet Propulsion Laboratory, California Institute of Technology under a contract with NASA. Support for this work was provided by NASA through an award issued by JPL/Caltech.
Some of the data presented herein were obtained at Palomar Observatory, which is operated by a collaboration between California Institute of Technology, Jet Propulsion Laboratory, Yale University, and National Astronomical Observatories of China. 
Some of the data presented herein were obtained at the W. M. Keck Observatory, which is operated as a scientific partnership among the California Institute of Technology, the University of California and the National Aeronautics and Space Administration. The Observatory was made possible by the generous financial support of the W. M. Keck Foundation. The authors wish to recognize and acknowledge the very significant cultural role and reverence that the summit of Maunakea has always had within the indigenous Hawaiian community.  We are fortunate to have the opportunity to conduct observations from this mountain.

\textit{Facilities}: Spitzer (IRAC), Hale (WIRC),  Magellan:Baade (FourStar), Keck:I (MOSFIRE), Keck:II (NIRC2), Swift (XRT)

\onecolumngrid

\clearpage

\begin{LongTables}
\begin{deluxetable}{lllllllll}
\tablecaption{ULX Candidate Sample}
\tablewidth{0pt}
\tablehead{Galaxy & SPIRITS &ULX Name & RA & DEC & Error & Distance & Dist. & Pos Ref \\
& Gal (Y/N)& & (J2000) & (J2000) & ('') & (Mpc) &  Ref & }
\startdata
NGC 300      & Y & SN~2010da/NGC 300 ULX1        & 00:55:04.93  & -37:41:43.9   & 0.9 & 1.88$\pm$0.05  & A  & Binder et al.~(2016) \\
IC 342      & Y & {[}LB2005{]} PGC 13826 ULX2        & 03:46:45.4  & 68:09:47.3   & 1.1 & 2.73$\pm$0.19  & B  & L17                               \\
IC 342      & Y & CXO J034657.4+680619               & 03:46:57.4  & 68:06:19.1   & 0.6 & 2.73$\pm$0.19  & B  & Evans et al. (2010)               \\
IC 342       & Y & XMMU J034555.6+680455              & 03:45:55.61 & 68:04:55.3   & 0.3 & 2.73$\pm$0.19   & B  & Feng \& Kaaret (2008)             \\
IC 342       & Y & {[}SST2011{]} J034615.64+681112.2  & 03:46:15.61 & 68:11:12.8   & 0.4 & 2.73$\pm$0.19   & B & Lin et al.~(2012)                 \\
NGC 4190    & N & CXO J121345.2+363754               & 12:13:45.3  & 36:37:54.7   & 0.9 & 2.83$\pm$0.28  & B  & Evans et al.~(2010)               \\
NGC 2403    & Y & CXOU J073625.6+653539              & 07:36:25.55 & 65:35:40.0   & 0.7 & 3.18$\pm$0.19  & B  & Swartz et al. (2004)              \\
NGC 2403    & Y & 2E 0732.2+6546                     & 07:37:02.33 & 65:39:35.0   & 0.7 & 3.18$\pm$0.19  & B  & H14                               \\
\enddata
\tablecomments{Properties of the ULX sample studied in this work including the host galaxy, whether or not it is included in the SPIRITS survey, ULX name (*name used in this work), RA and Declination of the X-ray position, 2-$\sigma$ X-ray positional uncertainty, adopted distance to the ULX, distance reference, and X-ray position reference. The distance references are as follows: (A) Gieren et al.~(2005), (B) Tully et al.~(2013),(C) Dalcanton et al.~(2009), (D) Pietrzynski et al.~(2010), (E) McQuinn et al.~(2010),(F) Tully et al.~(2009), (G) Jacobs et al.~(2009) (H) Karachentsev et al.~(2003), (I) Tully (1988),  (J) Shappee \& Stanek~(2011), (K) Drozdovsky \& Karachentsev (2000), (L) Poznanski et al.~(2009), (M) Tikhonov, Lebedev \& Galazutdinova (2015), (N) Springob et al.~(2009), (O) Olivares E. et al.~(2010), (P) Willick et al.~(1997)}
\label{tab:ULX}
\end{deluxetable}
\end{LongTables}

\clearpage

\begin{LongTables}
\begin{deluxetable}{lllllllll}
\tablecaption{\spitzer/IRAC Observations of Candidate Stellar Mid-IR ULX Counterparts}
\tablewidth{0pt}
\tablehead{Galaxy & ULX Name & MJD & Chnl & Flux ($\mu$Jy) & Mag & Abs Mag & PI & Prog ID}
\startdata
NGC 300 & NGC 300 ULX1 & 52964.06 & 1 & 120.6$\pm$2.46 & 15.92$\pm$0.02 & -10.45$\pm$0.02$\pm$0.06 & Helou & 1083 \\
NGC 300 & NGC 300 ULX1 & 54462.88 & 1 & 122.24$\pm$2.66 & 15.9$\pm$0.02 & -10.47$\pm$0.02$\pm$0.06 & Kennicutt & 40204 \\
NGC 300 & NGC 300 ULX1 & 54462.94 & 1 & 116.16$\pm$2.73 & 15.96$\pm$0.03 & -10.41$\pm$0.03$\pm$0.06 & Kennicutt & 40204 \\
NGC 300 & NGC 300 ULX1 & 55186.87 & 1 & 171.71$\pm$3.02 & 15.53$\pm$0.02 & -10.84$\pm$0.02$\pm$0.06 & Freedman & 61002 \\
NGC 300 & NGC 300 ULX1 & 55209.69 & 1 & 277.25$\pm$2.72 & 15.01$\pm$0.01 & -11.36$\pm$0.01$\pm$0.06 & Freedman & 61002 \\
NGC 300 & NGC 300 ULX1 & 55404.18 & 1 & 274.77$\pm$3.05 & 15.02$\pm$0.01 & -11.35$\pm$0.01$\pm$0.06 & Freedman & 61002 \\
NGC 300 & NGC 300 ULX1 & 55424.1 & 1 & 222.33$\pm$2.96 & 15.25$\pm$0.01 & -11.12$\pm$0.01$\pm$0.06 & Freedman & 61002 \\
NGC 300 & NGC 300 ULX1 & 55439.76 & 1 & 192.33$\pm$2.62 & 15.41$\pm$0.01 & -10.96$\pm$0.01$\pm$0.06 & Freedman & 61002 \\
\enddata
\tablecomments{\spitzer/IRAC observation summary of the stellar mid-IR ULX counterpart candidates including the host galaxy, ULX name, date of observation, IRAC channel, measured flux, apparent magnitude (in Vega), absolute magnitude, PI of observation, and Program ID of observation.}
\label{tab:ULXDetTab}
\end{deluxetable}
\end{LongTables}
\clearpage

\begin{turnpage}
\begin{deluxetable}{llllllllll}
\tablecaption{Median \spitzer/IRAC Flux from Candidate Stellar Mid-IR ULX Counterparts}
\tablewidth{0pt}
\tablehead{Galaxy & ULX Name & IR Class & Med Ch1 Flux & Med Ch2 Flux & Med Ch1 Mag & Med Ch2 Mag & Med Ch1 AbsMag & Med Ch2 AbsMag & Color \\}
\startdata
NGC 300 & NGC 300 ULX1 (Pre-2010) & Red & 120.6$\pm$3.54 & 135.82$\pm$2.86 & 15.92$\pm$0.03 & 15.3$\pm$0.02 & -10.45$\pm$0.07 & -11.07$\pm$0.06 & 0.62 $\pm$ 0.04 \\
NGC 300 & NGC 300 ULX1 (Post-2010) & Color Gap & 212.13$\pm$3.42 & 200.35$\pm$2.23 & 15.3$\pm$0.02 & 14.88$\pm$0.01 & -11.07$\pm$0.06 & -11.49$\pm$0.06 & 0.42 $\pm$ 0.02 \\
Holmberg II & Holmberg II X-1 & Red & 22.45$\pm$2.27 & 30.38$\pm$2.3 & 17.74$\pm$0.11 & 16.93$\pm$0.08 & -9.91$\pm$0.11 & -10.72$\pm$0.09 & 0.81 $\pm$ 0.14 \\
NGC 7793 & NGC7793 P13 & -* & 5.71$\pm$1.42 & 3.52$\pm$0.98 & 19.23$\pm$0.27 & 19.27$\pm$0.3 & -8.46$\pm$0.29 & -8.42$\pm$0.31 & -0.04 $\pm$ 0.4 \\
NGC 253 & NGC 253 ULX1 & Blue & 23.97$\pm$5.27 & $>$12.28 & 17.67$\pm$0.24 & $>$17.91 & -10.03$\pm$0.26 & $<$-9.79 & $<$-0.24\\
Holmberg IX & Holmberg IX X-1 & -* & 3.2$\pm$0.54 & 3.18$\pm$0.67 & 19.86$\pm$0.18 & 19.38$\pm$0.23 & -7.93$\pm$0.2 & -8.41$\pm$0.24 & 0.48 $\pm$ 0.29 \\
M81 & NGC 3031 ULX1 & Color Gap & 11.53$\pm$4.01 & 10.73$\pm$2.76 & 18.47$\pm$0.38 & 18.06$\pm$0.28 & -9.32$\pm$0.4 & -9.73$\pm$0.31 & 0.41 $\pm$ 0.47 \\
M101 & M101 XMM1 & Blue & 19.84$\pm$3.35 & 14.07$\pm$2.32 & 17.88$\pm$0.18 & 17.77$\pm$0.18 & -11.16$\pm$0.19 & -11.27$\pm$0.19 & 0.11 $\pm$ 0.25 \\
\enddata
\tablecomments{Properties of the stellar mid-IR ULX counterpart candidates including the host galaxy, ULX name, IR color classification, median Ch1 and Ch2 flux (in $\mu$mJy, apparent mag, and absolute mag), and $[3.6]-[4.5]$ color. *IR color classifications are not provided for NGC7793 P13 and Holmberg IX X-1 whose mid-IR emission arises from a late-B supergiant mass donor (Motch et al.~2014) and variable jet (this work, Sec.~\ref{Sec:HoIX}), respectively.}
\label{tab:ULXDetColorTab}
\end{deluxetable}
\end{turnpage}
\clearpage

\begin{turnpage}
\begin{deluxetable}{lllllllll}
\tablecaption{Median \spitzer/IRAC Flux from Candidate star cluster, AGN, and background galaxy Mid-IR ULX Counterparts}
\tablewidth{0pt}
\tablehead{Galaxy & ULX Name & NIR/MIR Class & Med Ch1 AbsMag & Med Ch2 AbsMag & Med Ch3 AbsMag & Med Ch4 AbsMag & Ch1 - Ch2 & Ch3 - Ch4}
\startdata
Holmberg I & 1WGA J0940.0+7106 & AGN/AGN & -  & -13.0$\pm$0.029 &  - & -14.62$\pm$0.034 & -   &  -
\\
M101 & [LB2005] NGC 5457 X26 & SC/Hypernova Remnant & -13.49$\pm$0.174 & -14.21$\pm$0.141 & -14.47$\pm$0.234 & -15.88$\pm$0.216 & 0.72$\pm$0.212 & 1.41$\pm$0.31 \\
M101 & 2XMM J140248.0+541350 & AGN/AGN & -13.26$\pm$0.053 & -13.81$\pm$0.053 & -14.66$\pm$0.067 & -16.17$\pm$0.054 & 0.55$\pm$0.023 & 1.51$\pm$0.048 \\
M101 & [LB2005] NGC 5457 X32 & AGN/AGN & -14.44$\pm$0.051 & -14.49$\pm$0.051 & -14.67$\pm$0.07 & -15.4$\pm$0.065 & 0.05$\pm$0.011 & 0.73$\pm$0.064 \\
NGC 4258 & RX J121845.6+472420 & U/SC & -12.37$\pm$0.113 & -12.77$\pm$0.114 & -13.73$\pm$0.133 & -15.12$\pm$0.124 & 0.4$\pm$0.041 & 1.39$\pm$0.095 \\
NGC 4258 & RX J121844.0+471730 & SC/SC & -13.58$\pm$0.13 & -13.95$\pm$0.123 & -15.4$\pm$0.145 & -17.05$\pm$0.147 & 0.37$\pm$0.089 & 1.65$\pm$0.136 \\
NGC 4258 & RX J121857.7+471558 & U/SC & -13.52$\pm$0.128 & -13.73$\pm$0.127 & -16.49$\pm$0.118 & -18.27$\pm$0.118 & 0.21$\pm$0.092 & 1.78$\pm$0.06 \\
NGC 4258 & [LB2005] NGC 4258 X9 & AGN/AGN & -15.83$\pm$0.11 & -16.5$\pm$0.11 & -  &  - & 0.67$\pm$0.004 &  -  \\
\enddata
\tablecomments{Properties of the star cluster or background AGN mid-IR ULX counterpart candidates including the apparent host galaxy, ULX name, NIR (H14 and/or L17)/ MIR classification, median Ch1, Ch2, Ch3, and Ch4 absolute mag), $[3.6]-[4.5]$, and $[5.8]-[8.0]$ colors.}
\label{tab:ULXODetColorTab}
\end{deluxetable}
\end{turnpage}
\clearpage

\begin{deluxetable}{lllllll}
\tablecaption{Near-IR Observations of Candidate Stellar Mid-IR ULX Counterparts}
\tablewidth{0pt}
\tablehead{Galaxy & ULX Name & MJD & Instrument & Filter & Mag & Abs Mag}
\startdata
Holmberg II & Holmberg II X-1 & 58126.40 & MOSFIRE & Ks & 19.51$\pm$0.16 & -8.14$\pm$0.16$\pm$0.03 \\
Holmberg II & Holmberg II X-1 & 58126.41 & MOSFIRE & H & 20.21$\pm$0.35 & -7.44$\pm$0.35$\pm$0.03 \\
Holmberg II & Holmberg II X-1 & 58450.61 & NIRC2-LGS & Ks (1.0'') & 19.81$\pm0.05$  & -7.70$\pm$0.05$\pm$0.03  \\
 &  & & NIRC2-LGS & Ks (0.2'') & 20.50$\pm0.1$ & -7.0$\pm$0.1$\pm$0.03 \\
Holmberg II & Holmberg II X-1 & 58450.54 & NIRC2-LGS & H (1.0'') & 20.42$\pm$0.06 & -7.08$\pm$0.06$\pm$0.03  \\
 &  & & NIRC2-LGS & H (0.2'') & 20.84$\pm$0.1 & -6.66$\pm$0.1$\pm$0.03 \\
M81 & NGC 3031 ULX1 & 	58184.36 & WIRC & Ks & $>$18.85 & $>$-8.94 \\
M81 & NGC 3031 ULX1 & 58185.93 & WIRC & H & $>$20.12 & $>$-7.67 \\
\enddata
\tablecomments{Properties of the near-IR observations of the stellar mid-IR ULX counterpart candidates including the ULX host galaxy, ULX name, date of observations, instrument, imaging filter band, apparent magnitude, and absolute magnitude. For the NIRC2-LGS observations, the photometry measured with $1''$ and $0.2''$ radius apertures are shown and indicated in the ``Filter" field.}
\label{tab:ULXNIRTab}
\end{deluxetable}

\clearpage

\begin{deluxetable}{lllllllll}
\tablecaption{Candidate Star Clusters, AGN, and Background Galaxy ULX Counterparts}
\tablewidth{0pt}
\tablehead{Galaxy & ULX Name & MJD & Chnl & Flux ($\mu$Jy) & Mag & Abs Mag & PI & Prog ID}
\startdata
Holmberg I & 1WGA J0940.0+7106 & 53308.7 & 2 & 185.79$\pm$0.88 & 14.96$\pm$0.005 & -12.99$\pm$0.005$\pm$0.03 & Kennicutt & 159 \\
M101 & 2XMM J140248.0+541350 & 55960.72 & 1 & 142.75$\pm$1.69 & 15.73$\pm$0.013 & -13.31$\pm$0.013$\pm$0.05 & Kasliwal & 80196 \\
M101 & 2XMM J140248.0+541350 & 55960.72 & 2 & 140.11$\pm$1.22 & 15.27$\pm$0.01 & -13.77$\pm$0.01$\pm$0.05 & Kasliwal & 80196 \\
M101 & [LB2005] NGC 5457 X26 & 53072.49 & 1 & 164.43$\pm$19.13 & 15.58$\pm$0.127 & -13.46$\pm$0.127$\pm$0.05 & Rieke & 60 \\
M101 & [LB2005] NGC 5457 X26 & 57247.82 & 2 & 219.63$\pm$18.65 & 14.78$\pm$0.093 & -14.26$\pm$0.093$\pm$0.05 & Kasliwal & 11063 \\
M101 & [LB2005] NGC 5457 X32 & 57635.4 & 1 & 404.23$\pm$1.0 & 14.6$\pm$0.003 & -14.44$\pm$0.003$\pm$0.05 & Kasliwal & 11063 \\
M101 & [LB2005] NGC 5457 X32 & 57845.07 & 2 & 280.23$\pm$0.74 & 14.52$\pm$0.003 & -14.52$\pm$0.003$\pm$0.05 & Kasliwal & 13053 \\
NGC 4258 & [LB2005] NGC 4258 X9 & 55210.6 & 1 & 1126.04$\pm$2.55 & 13.49$\pm$0.002 & -15.83$\pm$0.002$\pm$0.11 & Freedman & 61008 \\
\enddata
\tablecomments{Ch1 and Ch2 \spitzer/IRAC observation summary of the star cluster, AGN, or background galaxy mid-IR ULX counterpart candidates including the host galaxy, ULX name, date of observation, IRAC channel, measured flux, apparent magnitude (in Vega), absolute magnitude, PI of observation, and Program ID of observation. The selected \spitzer/IRAC were chosen based on the closest timing to the near-IR observation by H14 and L17. }
\label{tab:ULXTabOD}
\end{deluxetable}

\clearpage

\begin{LongTables}
\begin{deluxetable}{lllllllll}
\tablecaption{\spitzer/IRAC Mid-IR Non-Detections of ULXs}
\tablewidth{0pt}
\tablehead{Galaxy & ULX Name & MJD & Chnl & Flux ($\mu$Jy) & Mag & Abs Mag & PI & Prog ID}
\startdata
IC 342 & J034657.4+680619 & 56810.6 & 2 & $<$31.16 & $>$16.68 & $>$-10.51 & Kasliwal & 10136 \\
IC 342 & J034657.4+680619 & 56810.6 & 1 & $<$29.09 & $>$17.25 & $>$-9.93 & Kasliwal & 10136 \\
IC 342 & J034657.4+680619 & 56979.87 & 1 & $<$22.47 & $>$17.53 & $>$-9.65 & Kasliwal & 10136 \\
IC 342 & J034657.4+680619 & 56979.87 & 2 & $<$14.22 & $>$17.53 & $>$-9.65 & Kasliwal & 10136 \\
IC 342 & PGC 13826 ULX2 & 56810.6 & 1 & $<$18.3 & $>$17.75 & $>$-9.43 & Kasliwal & 10136 \\
IC 342 & PGC 13826 ULX2 & 56810.6 & 2 & $<$17.73 & $>$17.29 & $>$-9.89 & Kasliwal & 10136 \\
IC 342 & PGC 13826 ULX2 & 56979.87 & 2 & $<$37.47 & $>$16.47 & $>$-10.71 & Kasliwal & 10136 \\
IC 342 & PGC 13826 ULX2 & 56979.87 & 1 & $<$28.53 & $>$17.27 & $>$-9.91 & Kasliwal & 10136 \\
\enddata
\tablecomments{Summary of the non-detections of mid-IR ULX counterparts with \spitzer/IRAC including the ULX host galaxy, ULX name, date of observation, IRAC channel, 3-$\sigma$ upper limit of detection (in $\mu$Jy, apparent magnitude, and absolute magnitude), PI of the observation, and program ID of the observation.}
\label{tab:ULXNonDetTab}
\end{deluxetable}
\end{LongTables}
\clearpage


\begin{deluxetable}{lllllll}
\tablecaption{``Red" ULX Dust Properties}
\tablewidth{0pt}
\tablehead{ULX Name & $L_X$ (erg s$^{-1}$) & $T_d$ (K) & $L_{IR}$ ($10^4$ L$_\odot$) & $M_d$ ($10^{-6}$ M$_\odot$) & $R_\mathrm{eq}^{IR}$ (AU) & $R_\mathrm{sub}$ (AU)}
\startdata
Holmberg II X-1 & $7.7\times10^{39}$ & $720^{+140}_{-100} $& $2.6^{+0.8}_{-0.4}$ & $1.1^{+1.2}_{-0.6}$ & $  130$ & 40 \\
NGC 925 ULX1 & $3\times10^{40}$ & $730^{+130}_{-100} $& $9.5^{+2.6}_{-1.4}$ & $3.8^{+3.9}_{-1.9}$ & $  240$ & 70\\
NGC 4631 X4 & $2\times10^{39}$ & $820^{+310}_{-180} $& $16.8^{+6.8}_{-0.4}$ & $4.8^{+9.8}_{-3.2}$ & $  260$ & 20\\
NGC 300 ULX1 (Pre-2010) & $\sim2\times10^{39}$ & $870^{+50}_{-50} $& $3.0^{+0.1}_{-0.1}$ & $0.7^{+0.1}_{-0.2}$ & $  100$ & 20
\enddata
\tablecomments{X-ray, IR, and derived dust properties of the 4 ``red" mid-IR ULX counterpart candidates including the ULX name, X-ray (2 - 10 keV) luminosity, dust temperature, IR luminosity, total dust mass, dust equilibrium temperature radius, dust sublimation radius. A heating source with a luminosity equivalent to the IR luminosity and a 20000 K photosphere ($\left<Q\right>_{*}/ \left<Q\right>_{IR} = 30$) is adopted for equilibrium radius calculation. For the R$_\mathrm{sub}$ calculation, a heating source with a luminosity equivalent to $L_X$ value is adopted and $\left<Q\right>_{X}/ \left<Q\right>_{IR} = 1$.}
\label{tab:ULXDustTab}
\end{deluxetable}
\clearpage

\end{document}